\documentclass[preprintnumbers,unsortedaddress,superscriptaddress,notitlepage,nofootinbib]{revtex4-1}
\usepackage{amsfonts}
\usepackage{mathrsfs}
\usepackage{amsmath}
\usepackage{mathrsfs}
\usepackage{amssymb}
\usepackage[english]{babel}
\usepackage{graphicx}
\usepackage{simplewick}
\usepackage{color}
\usepackage{soul}
\usepackage{caption}
\usepackage{subcaption}
\usepackage{booktabs}
\usepackage[utf8]{inputenc}

\begin{document}

\title{Non-Abelian family symmetries as portals to dark matter}
\author{I. de Medeiros Varzielas}
\email{ivo.de@soton.ac.uk}
\affiliation{{\small School of Physics and Astronomy, University of Southampton,}\\
Southampton, SO17 1BJ, U.K.}

\author{O. Fischer}
\email{oliver.fischer@unibas.ch}
\affiliation{{\small Department of Physics, University of Basel,}\\
Klingelbergstr. 82, CH-4056 Basel, Switzerland}

\begin{abstract}
Non-Abelian family symmetries offer a very promising explanation for the flavour structure in the Standard Model and its extensions.
We explore the possibility that dark matter consists in fermions that transform under a family symmetry, such that the visible and dark sector are linked by the familons - Standard Model gauge singlet scalars, responsible for spontaneously breaking the family symmetry.
We study three representative models with non-Abelian family symmetries that have been shown capable to explain the masses and mixing of the Standard Model fermions. 

One of our central results is the possibility to have dark matter fermions and at least one familon with masses on and even below the experimentally accessible TeV scale.
In particular we discuss the characteristic signatures in collider experiments from light familon fields with a non-Abelian family symmetry, and we show that run I of the LHC is already testing this class of models.
\end{abstract}

\maketitle

\section{Introduction}
The Standard Model (SM) is a very efficient and well tested theory that explains almost all the experimental observations in elementary particle physics.
However, there are several observations that the SM does not address and which motivate the extensions Beyond the SM (BSM).
Two prominent examples of unresolved observations are the flavour problem, and the observation of collisionless, cold Dark Matter (DM).
The fact that the fermions are arranged in three generations has been addressed recently by ref.~\cite{vanderBij:2010nu} with an argument that connects the SM gauge group with the number of fermion generations. However, the puzzling pattern of hierarchical masses and small fermion mixing in the quark sector, opposed to the large mixing in the lepton sector, is beyond present understanding.
Furthermore, the SM does not offer a suitable candidate that can take the r\^{o}le of the DM to make up the observed relic density that is about five times more abundant than baryonic matter \cite{Ade:2013zuv}.

The experimental status of DM signals are not conclusive: there are some hints for DM signals in indirect detection experiments, but there is no unambiguous signal in direct detection experiments, see e.g.\ refs.~\cite{Yuan:2014rca,Gomez-Vargas:2014gra,Calore:2015nua,O'Leary:2015gfa}.
From a theoretical point of view, a new unbroken symmetry can be used to explain the DM properties by preventing decays of the DM into SM particles.
So-called ``top-down'' approaches often generate a ``dark sector'', as is the case e.g.\ for Supersymmetry (SUSY) with $R$ parity, extra dimensions and Grand Unified Theories (GUTs). In the above mentioned examples, at least one new symmetry is introduced, which distinguishes a new sector from the SM. 
Symmetries are also the most promising BSM approach to deal with the flavour problem, i.e.\ the peculiar structure exhibited by the masses and mixings of the fermions. In this approach, the symmetry is often referred to as a Family Symmetry (FS), as it relates the generations of fermions, see ref.~\cite{King:2014nza} for a recent review.

FSs are particularly appealing in BSM theories with additional particles, such as additional Higgs fields \cite{Varzielas:2011jr}, as the FS will control the flavour problem, while the theory also addresses other shortcomings of the SM, for instance by providing suitable DM candidates \cite{Varzielas:2015joa}.

More often than not, the introduced DM candidate is an additional particle (scalar or fermion) charged under an Abelian symmetry. Usually the simplest possibility of a $Z_2$ symmetry is evoked, because the DM phenomenology does not depend very much on the explicit choice for the stabilising symmetry, as long as it prevents the decay of the lightest particle charged under it.
The protective symmetry for the DM leads to having two different symmetries at the heart of the theory: the non-Abelian (gauge) symmetry in the SM sector, and the symmetry in the DM sector. 
The most often studied interaction between the dark and the visible sector is the so-called ``Higgs portal'', where the dark and visible sector interact via the Higgs boson, and which leads to the SM final states from DM annihilation to be spread over all the kinematically available SM particle content according to their Yukawa couplings.

Recently, the possibility of a deeper connection between the DM and flavour were studied in different frameworks by independent groups.
The generic properties of flavoured DM -- the Flavon or Familon portal -- were investigated for an Abelian FS \cite{Calibbi:2015sfa}, where it was shown that, compared to the Higgs portal, the DM in the Familon portal setup allows for smaller DM masses since the direct detection constraints lose some of their predictivity. 
A very detailed study investigated the connection of flavoured (fermionic or scalar) DM and the SM to be given by a gauged FS, where the Yukawa matrices are promoted into physical scalar fields \cite{Bishara:2015mha}. The authors found that the DM and FS mass scale should be $\sim$TeV to match the observational constraints.
Furthermore, in ref.\ \cite{Chen:2015jkt} a minimal flavor violation scenario was compared with an $SU(3)$ FS, where the interaction between SM and DM is mediated by a FS singlet.

In this paper, we study the phenomenological implications of effective models with non-Abelian FS and flavoured DM.
We consider non-Abelian FSs because the observed leptonic mixing shows hints that those are preferred when fitting the observed patterns \cite{Ma:2004zd, deMedeirosVarzielas:2005ax, deMedeirosVarzielas:2011tp}. Typically the FS is broken by familon fields developing a vacuum expectation value (VEV), which can explain the observed flavour structure.
For fermionic DM particles that are singlets under the SM gauge group and reside in non-trivial multiplets of the FS, the interaction between visible and dark sector can be mediated exclusively by the {\it familons}, which are also often referred to as {\it flavons} in the literature.

The models considered here share the generic feature of a dark sector, that is charged under non-Abelian FS, together with scalar familon fields.
They are derived from realistic UV-complete FS models in a SUSY (and in one case a GUT) framework, and they are simplified such that the number of free model parameters can be related to observational constraints.
In the following we show, that with familons and DM masses around the electroweak scale, the observed relic abundance as well as experimental constraints on flavour changing neutral currents can be accommodated.
Furtherore we show, that specific signatures and predictions for collider phenomenology are possible, and that the present experimental (non-)observation of deviations from the SM predictions is already putting bounds on combinations of the model parameters.

The FS models are expected to be representative for the class of models where the interactions between dark and SM sector are exclusively given by the non-Abelian familon portal -- much in the same way that more generic SUSY models make similar predictions in terms of the dark sector compared to the MSSM.
They are not intended to explain the SM phenomenology in the fermion sector, but rather to capture the phenomenological features of the non-Abelian familon portal to the here considered dark sector.

We note that an interesting and new feature of the Familon portal that we explore in this article is the possibility that the underlying FS can give rise to observational features, that may allow to distinguish one FS from another one (and in particular determine if the FS is non-Abelian).
We therefore aim to highlight generic features of this type of DM.
In terms of discriminating power, we investigate observables that may help to distinguish different models, e.g.\ whether the familon is coupling both to quarks and leptons (as in GUTs), as opposed to the familon coupling exclusively to the leptons.
Regarding the latter, we further distinguish the cases of the familon coupling either to the charged, or the neutral lepton sector.

The structure of our paper is as follows. In section \ref{sec:DMstudy} we introduce three effective models with fermionic DM candidates and a different non-Abelian Familon portal.
In the same section, the relic abundance constraint as well as relevant other constraints are briefly studied. In section \ref{sec:colliderpheno}, we discuss the production of familons at lepton and hadron colliders, and the constraints resulting from a selection of present experimental observations.
In section \ref{sec:conclusions} we summarise our results and, after a brief discussion, present our conclusions.

\section{Specific models}
\label{sec:DMstudy}
In FS models, the SM fermions are usually embedded in a specific representation of the FS, and extra scalar SM gauge singlet fields -- the familons -- are introduced, that acquire VEVs with a specific alignment, such that the FS is broken and the fermion masses and mixings can be explained. 

In this section, we present three different flavour models based on UV-complete theories, that can lead to a realistic description of the pattern of fermions' masses and mixings. The models share the general feature of additional SM gauge singlet fermions, that are introduced as DM candidates and arranged as a multiplet of the respective FS, such that it exclusively couples to the familon fields.

Two of the models are based on the $A_4$ symmetry, the group multiplication of which is listed in Appendix \ref{app:A4}, for the chosen basis.
The models are based on the UV complete models in a SUSY framework \cite{Varzielas:2010mp} that account for the observed value of $\theta_{13}$ \cite{Varzielas:2012ai}.
Therein, the SM leptons are embedded into a global $A_4$ symmetry and the experimentally observed leptonic mixing is generated through two $A_4$-triplet scalars $\phi_\ell$ and $\phi_\nu$, and exclusive couplings to the charged and neutral leptons, respectively. Their specific VEVs are aligned through the so-called {\it F-term alignment} within the SUSY framework.
In this type of model it is possible to extend the $A_4$ to address also the quark sector. However, for the sake of simplicity, we choose to keep the quark fields as trivial $A_4$-singlets, which allows to fit their masses and mixings as in the SM.

The third model of this section considers a FS embedding in a SUSY GUT framework as in ref.~\cite{deMedeirosVarzielas:2005ax}, which is based on $SO(10)$ GUT and a continuous $SU(3)$ FS which necessarily addresses both quarks and leptons. Similar frameworks were explored in ref.\ \cite{deMedeirosVarzielas:2011wx,Varzielas:2012ss} wherein discrete subgroups were used in order to generate different paths to a non-zero reactor mixing angle $\theta_{13}$.

\subsection{$A_4$ lepton model with leptophilic DM}
\begin{table}
  \centering
  \begin{tabular}{c||cccc|ccccc|cc}
    \hline
     & $\chi$ & $\chi^c_1$ & $\chi^c_2$ & $\chi^c_3$ & $\nu^{c}$ & $L$ & $e^{c}$ & $\mu^{c}$ & $\tau^{c}$  & $H$ & $\phi_{\ell}$ \\
    \hline\hline
    $A_{4}$ & $3$ & $1$  & $1''$  & $1'$ & $3$  & $3$  & $1$  & $1''$  & $1'$  & $1$  & $1$   \\
$U(1)_{B-L}$ & $0$ & $0$ & $0$ & $0$ & $1$ & $-1$ & $1$ & $1$ & $1$ & $0$ &  $0$ \\  
    \hline
  \end{tabular}
   \caption{Field assignment within $A_4 \times U(1)_{B-L}$. New fields are gauge singlets with respect to the SM gauge group. The SM gauge charges for the SM fields are unchanged.}
  \label{tab:assignment1}
\end{table}

As mentioned above, we posit a global $A_4$ symmetry, and add to the field content an extra scalar field $\phi_\ell$ -- the familon -- that is a SM gauge singlet, and transforms non-trivially with respect to the $A_4$ symmetry.
The left-handed fields among the charged leptons of the SM field content are embedded into $A_4$-triplets and the right-handed leptons are assigned into the three different singlet representations of $A_4$.

The DM in this model consists in the $A_4$-triplet fermion field $\chi$ and the three $A_4$-singlet fermion fields $\chi^c_i$, $i=1,2,3$, all of which are singlets with respect to the SM gauge group.
In order to prevent a mixing between the $\chi^c_i$ with the right-handed neutrinos $\nu^c$, we further enforce the baryon-minus-lepton number to be a global symmetry: $U(1)_{B-L}$.
The field content of the model, along with the FS and $U(1)_{B-L}$ charges, is summarised in tab.~\ref{tab:assignment1}.
The Lagrangian density of the considered model is given by
\begin{equation}
\mathscr{L} = \mathscr{L}_\mathrm{SM} +  \mathscr{L}_\chi + \mathscr{L}_\phi \,,
\end{equation}
where $\mathscr{L}_\mathrm{SM}$ contains the usual SM field content (including their interactions with the familon), $\mathscr{L}_\chi$ includes the kinetic terms for the dark fermions as well as the Yukawa-like interactions with the familon, while $\mathscr{L}_\phi$ contains the terms of the scalar potential that include the familon field. 
We now turn to the discussion of the three contributing terms and their implications separately, although for $\mathscr{L}_\mathrm{SM}$ we limit the discussion to the charged lepton interactions with the familon fields.

\subsubsection{The scalar sector}
In the following we do not consider tri-linear combinations of the familon fields ($\propto [\phi_\ell \phi_\ell]_s \phi_\ell$), which is motivated because they are forbidden by additional symmetries present in the UV complete theory. We also disregard for now bi-linear combinations of familons (explicit mass-terms). We consider then the following non-SM part of the scalar potential:
\begin{align}
\mathscr{L}_\phi = & - c_{\phi,s} [\phi_\ell \phi_\ell^\dagger]_s [\phi_\ell \phi_\ell^\dagger]_s
- c_{1} [\phi_\ell \phi_\ell^\dagger]_{1} [\phi_\ell \phi_\ell^\dagger]_{1} - c_{2} [\phi_\ell \phi_\ell^\dagger]_{1^\prime} [\phi_\ell \phi_\ell^\dagger]_{1^{\prime\prime}} - c_{\phi_\ell H} [\phi_\ell \phi_\ell^\dagger] H^\dagger H \,,
\label{scalarpotential}
\end{align}
where the $H$ is the Higgs doublet. Note, that due to the $A_4$-product rules, the antisymmetric contractions of $\phi_\ell$-fields vanish identically and terms involving 
$[\phi_\ell \phi_\ell]_a$ were not shown above.

When the electroweak symmetry gets broken by the Higgs-VEV, $v_\mathrm{EW} =  246.22$ GeV, the coupling parameter $c_{\phi_\ell H}$ introduces the mass-term $c_{\phi_\ell H} v_\mathrm{EW}^2/2$ for the specific familon component $\phi_{\ell_1}$. For the coupling parameter being smaller than zero, the familon field itself acquires a vacuum expectation value (VEV) in a specific direction in flavour space:
\begin{align}
\langle \phi_l \rangle = (u,0,0)\,,
\end{align}
where the magnitude $u$ is given by
\begin{equation}
u = \frac{\sqrt{c_{\phi H}}\,v_\mathrm{EW}}{\sqrt{2 (c_{\phi,s} + 4\,c_1)}}\,.
\end{equation}
In the case of an approximate cancellation, $c_{\phi,s} + 4\,c_1 \simeq 0$, a hierarchy exists between the electroweak VEV and $u$: $v_\mathrm{EW} \ll u$. In the following, we assume that this hierarchy is present, since this is required in the UV-complete theory this model is based on.

After symmetry breaking, the familon field $\phi_\ell$ mixes with the real Higgs scalar $h$, with the mixing being proportional to the ratio of the VEVs, $v_\mathrm{EW}/u$, which is strongly suppressed due to the here considered VEV hierarchy. Considering only the coupling $c_{\phi,s}$ to be non-zero, the mass eigenstates are given by $\phi^0 = \phi_{\ell_1}$ and the linear combinations $\phi^\pm = \phi_{\ell_2}\pm\phi_{\ell_3}$, with the masses 
\begin{equation}
m_{\phi^0} = \sqrt{2\,c_{\phi,s}\,} u\,, \qquad m_{\phi^\pm} = \sqrt{c_{\phi,s}}\, u\,.
\end{equation}
It is worth noting, that the three resulting physical familons all reside on the same mass scale, defined by $u$.

\subsubsection{SM-familon interactions: Charged leptons}
The hierarchy of the charged lepton masses is explained by the  Froggatt-Nielsen (FN) mechanism \cite{Froggatt:1978nt}, and the specific renormalisable interactions between the charged leptons and the familons were considered in \cite{Varzielas:2010mp, Varzielas:2012ai}. When the heavy FN messenger fields are integrated out, the charge assignments from tab.~\ref{tab:assignment1} allow for the construction of an effective, non-renormalizable dimension five operator, that introduces an interaction between the SM charged leptons, the Higgs boson, and the familon
\begin{align}
O_{eff}^{d=5} \supset \frac{H}{\Lambda} \left( y_e [\phi_{\ell} L_i ] e^c + y_\mu [\phi_{\ell} L_i ]' \mu^c + y_\tau [\phi_{\ell} L_i ]'' \tau^c \right)\,,
\label{eq:Oeff5_ell}
\end{align}
where the $SU(2)_L$--doublet $H$ projects out the charged leptons from the $SU(2)_L$--doublets $L_i$ and $\Lambda$ encodes the mass scale and coupling of the FN messenger particle from the UV-complete theory, see \cite{Varzielas:2010mp, Varzielas:2012ai} for further details.
Expanding both, $\phi$ and $H$ around their respective VEVs yields from the invariant effective operators eq.~\eqref{eq:Oeff5_ell}  an effective Yukawa-like interaction between the familons and the charged leptons, and also between the Higgs boson and the charged leptons.
The mass matrix of the charged leptons arises after breaking of the electroweak and $A_4$ symmetries, when the VEVs of the scalar fields $\phi_{\ell_i}$ and $H$ are inserted:
\begin{equation}
  M_{L} \propto  \text{Diag}(y_e,y_\mu,y_\tau)\,.
\end{equation}
We emphasize, that Yukawa couplings of the physical charged leptons to the $\phi_{\ell_{2,3}}$ are purely off-diagonal, and can lead to interesting lepton flavour changing phenomena as was shown for cases with multiple Higgs bosons \cite{Heeck:2014qea,Varzielas:2015joa}.

\subsubsection{The dark sector}
With the $A_4$-triplet, $\chi$, and the three $A_4$-singlet fermions, $\chi^c_i,\,i=1,2,3$, the dark sector is mimicking the charged-lepton sector. Omitting the kinetic terms, the dark sector contains the following terms:
\begin{equation}
\mathscr{L}_\chi \supset - y_{\chi_1}\,[\phi_{\ell} \chi]\,\chi^c_1 - y_{\chi_2}\,[\phi_{\ell} \chi]^\prime\,\chi^c_2 - y_{\chi_3}\,[\phi_{\ell} \chi]^{\prime\prime}\,\chi^c_3
\end{equation}
After the $\phi_\ell$ develops its VEV, the following mass matrix of the $\chi$ fields emerges:
\begin{equation}
  M_{DM} = u \times \text{Diag}(y_{\chi_1},y_{\chi_2},y_{\chi_3})\,.
\end{equation}
The couplings of the physical familon $\phi^0$ (which consists exclusively of the component $\phi_{\ell_1}$) to the $\chi$ fields is also diagonal, as opposed to those of the lighter fields $\phi^\pm$ (which consist of the components $\phi_{\ell_{2,3}}$), that couple exclusively off-diagonally to the dark fermions.

The three parameters $y_{\chi_i},\,i=1,2,3$ allow for three different $\chi$-field masses. 
The $U(1)_{B-L}$--symmetry assignment to the fermions in tab.~\ref{tab:assignment1} prohibits direct couplings between the dark-sector fermions and SM leptons, such that decays of the former into the latter are forbidden, which in turn stablises the lightest dark fermion on cosmological time scales.

\subsubsection*{Dark Matter relic density}
\begin{figure}
\begin{center}\includegraphics[width=0.9\textwidth]{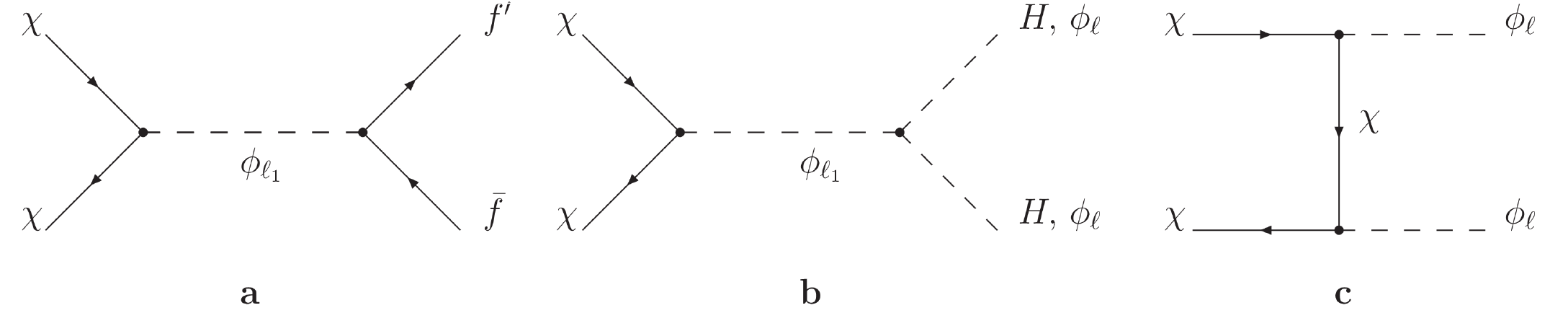}\end{center}
\caption{Feynman diagrams depicting the annihilation channels of the fermionic DM, $\chi$, which couples only to the (non-Abelian) familon portal $\phi_\ell$. {\bf a:} Annihilation via s-channel familon-exchange into two SM fermions (here, leptons only). {\bf b:} the same as {\bf a}, with the final state given by Higgs-scalars or familon fields, if kinematically allowed. {\bf c:} Annihilation via t-channel $\chi$-exchange into two familon fields. Note, that due to the possible decays of the familon into two (or more) SM fields, the familon is considered part of the SM field content.}
\label{fig:annihilationchannels}
\end{figure}
Before they freeze out from the primordial plasma, the DM, given by the $\chi$ fields, is kept in thermal equilibrium by the processes as depicted in fig.~\ref{fig:annihilationchannels}.
The details, i.e. which of the processes is dominant, are depending on the explicit relation of the masses $m_\chi$ and $m_\phi$. 
The amplitude from diagram \ref{fig:annihilationchannels} {\bf a} contributes to the total annihilation cross section of the DM candidates for all values of the model parameters, provided the couplings $y_{\chi_i}$ are non-zero. However, due to the small Yukawa couplings of the SM leptons, this amplitude alone is insufficient to give rise to the observed relic abundance.
The process corresponding to diagram \ref{fig:annihilationchannels} {\bf b}, on the other hand, is very powerful in creating a large annihilation cross section; a non-zero scalar coupling $c_{\phi H}$ can lead to the SM final states of two Higgs bosons (also counting the Goldstone bosons) or two familon fields (if kinematically available), or even a familon $\phi_{\ell_1}$ together with a Higgs boson. This potentially large contribution to the total annihilation cross section is very dependent on the mass relations of the familons, the DM fermions, and the Higgs boson mass.
In the case of the familon fields being (much) lighter than the DM fields, the contribution from the amplitude that corresponds to diagram \ref{fig:annihilationchannels} {\bf c} to the total annihilation cross section becomes dominant. Due to this process being a t-channel interaction, this makes the matching of the observed relic density possible for larger DM masses (not considering the special case of resonant annihilation, which is discussed below), compared to the other two processes.

\begin{figure}
\begin{center}
\includegraphics[width=0.45\textwidth]{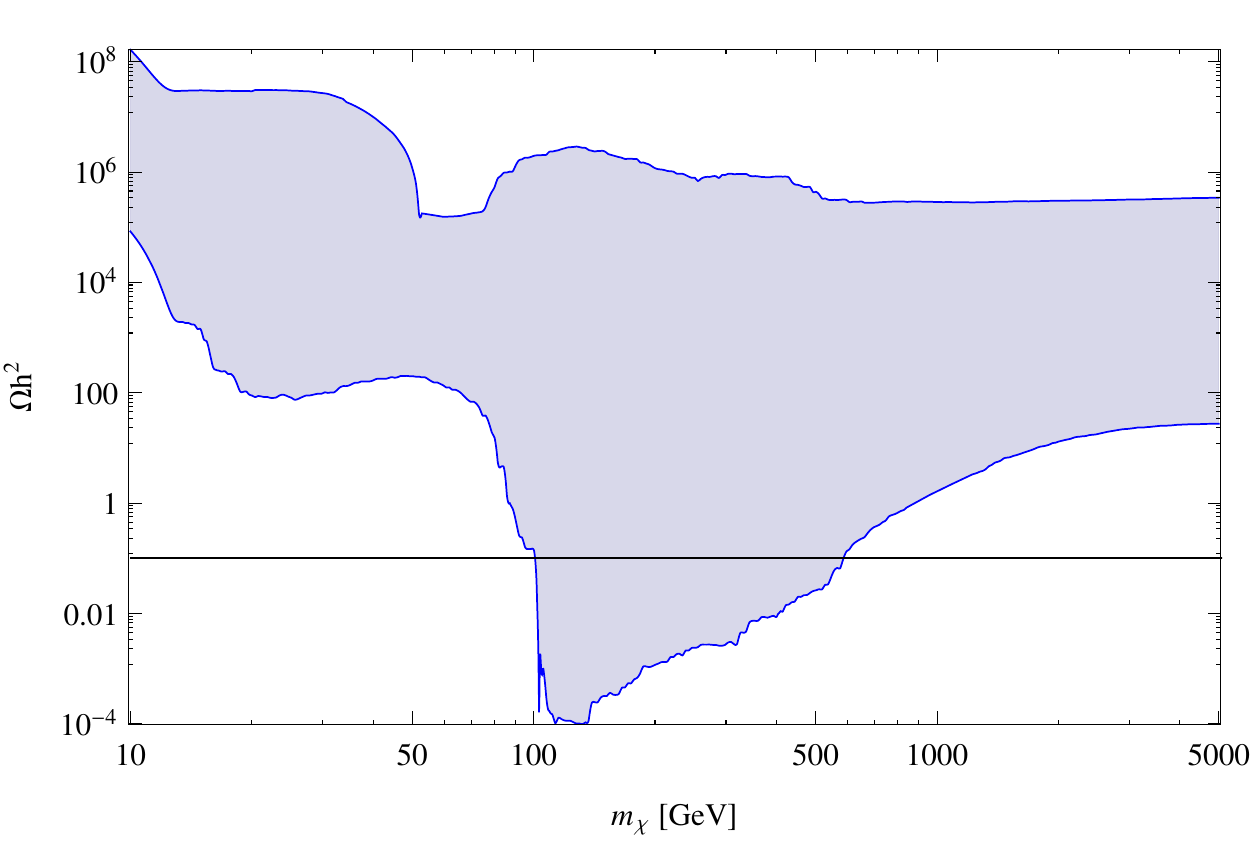}
\includegraphics[width=0.45\textwidth]{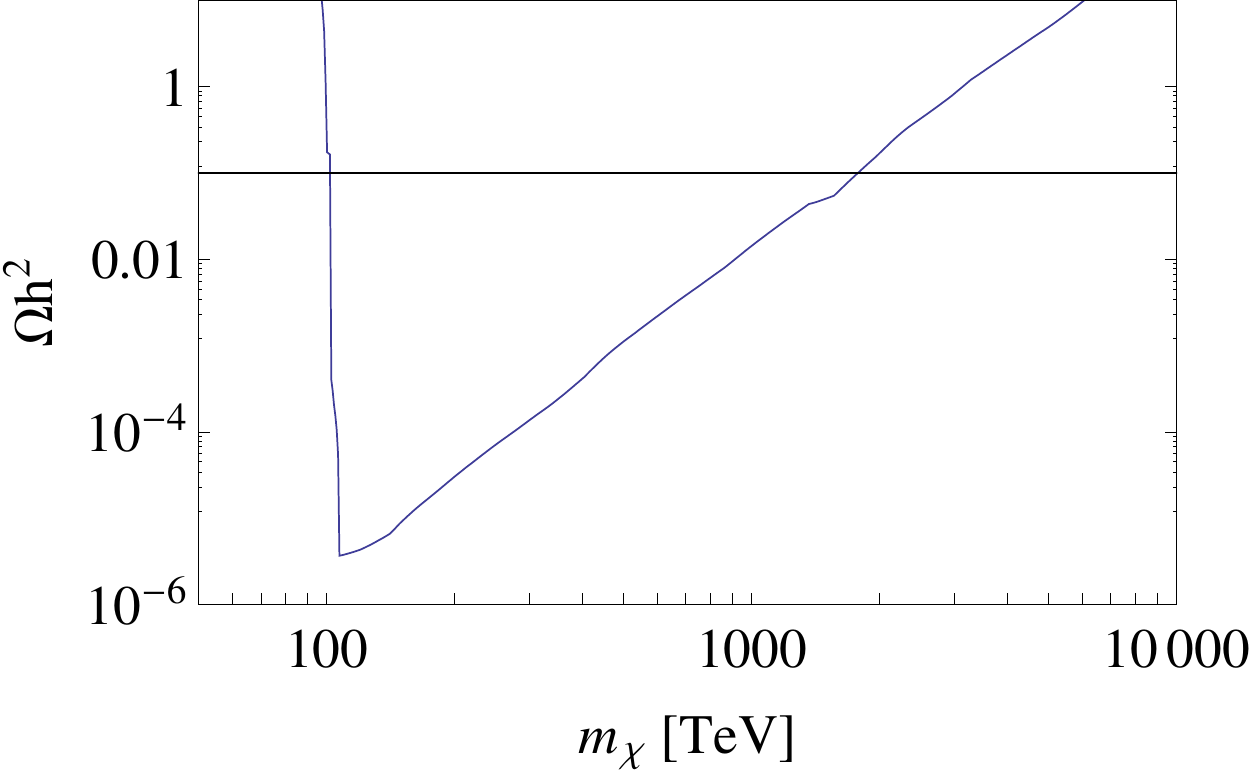}
\end{center}
\caption{{\it Left:} Overview of the produced relic density. The model parameters were marginalised within the numerical domains defined in eq.\ \eqref{eq:scatterscan-pars}. {\it Right:} Minimal relic densiy, imposing the parameters $y_{\chi i}= c_{\phi H} = 1$ and the relation $m_\chi \simeq 0.5\, m_{\phi^0}$ for resonant s-channel annihilation. Furthermore, $m_{\chi} := m_{\chi_1}$ and $m_{\chi_{2,3}} = 10$ TeV. The resulting relic density yields the maximum value for $m_\chi$, such that the observation can be matched.}
\label{fig:overview-phi-ell}
\end{figure}

The model as defined above has been implemented via {\tt Feynrules 2.3} \cite{Alloul:2013bka} into the numerical tool {\tt micrOMEGAs 4.1.8} \cite{Belanger:2014vza}. All the parameter space scans in the following have been carried out using the Tool for Parallel Processing in Parameter Scans ({\tt T3PS 1.0}) \cite{Maurer:2015gva}.
We show a general overview over the produced relic density $\Omega h^2$ in the left-hand plot of fig.~\ref{fig:overview-phi-ell}. Therein, we vary the parameters in the following ranges:
\begin{equation}
\begin{array}{c}
0.01 \leq y_{\chi_1}=y_{\chi_2}=y_{\chi_3},\,c_{\phi H} \leq 2 \\
10 \text{ GeV} \leq m_{\chi_1},m_{\chi_2},m_{\chi_3} \leq 10 \text{ TeV} \\
100 \text{ GeV} \leq m_{\phi_{2,3}} \leq 1 \text{ TeV} \\
m_{\phi_1} = \sqrt{2} m_{\phi_{2,3}}\\
u = \Lambda = 10 \text{ TeV}
\end{array}
\label{eq:scatterscan-pars}
\end{equation}
We note, that the possibility of resonant s-channel annihilation, i.e.\ the fine tuning of $m_\chi \simeq 0.5 m_{\phi_1}$ is excluded in this parameter scan.
The observational value for the DM relic density, being $\sim 0.1$ \cite{Ade:2013zuv}, is shown in the figure by the black line. 
It shows that the observation can be matched for DM masses, given by the minimum among the three $\chi_i$ field masses, between 100 GeV and $\sim$ 600 GeV, if the DM mass $m_{\chi}$ is smaller than the familon mass $m_{\phi_1}$.

Resonant s-channel annihilation occurs when the resonance condition is met:
\begin{equation}
m_\chi \sim 0.5\,m_\phi\,.
\label{eq:resonancecondition}
\end{equation}
This case, which can be considered as fine tuning of the model parameters, is a limiting one, as it leads to the maximal DM masses still compatible with the observed relic density. The right-hand plot of fig.~\ref{fig:overview-phi-ell} shows the lower contour of the relic density with resonant annihilation, and thus yields the upper bound on $m_\chi$, that is compatible with the observed relic density, to be $\sim 1.08$ TeV. For DM masses below this bound, the observed value for the relic density can be matched by adjusting (i.e. reducing) the coupling parameters $c_{\phi H}$ or $y_{\chi_i}$.

\begin{figure}
\begin{center}
\includegraphics[width=0.6\textwidth]{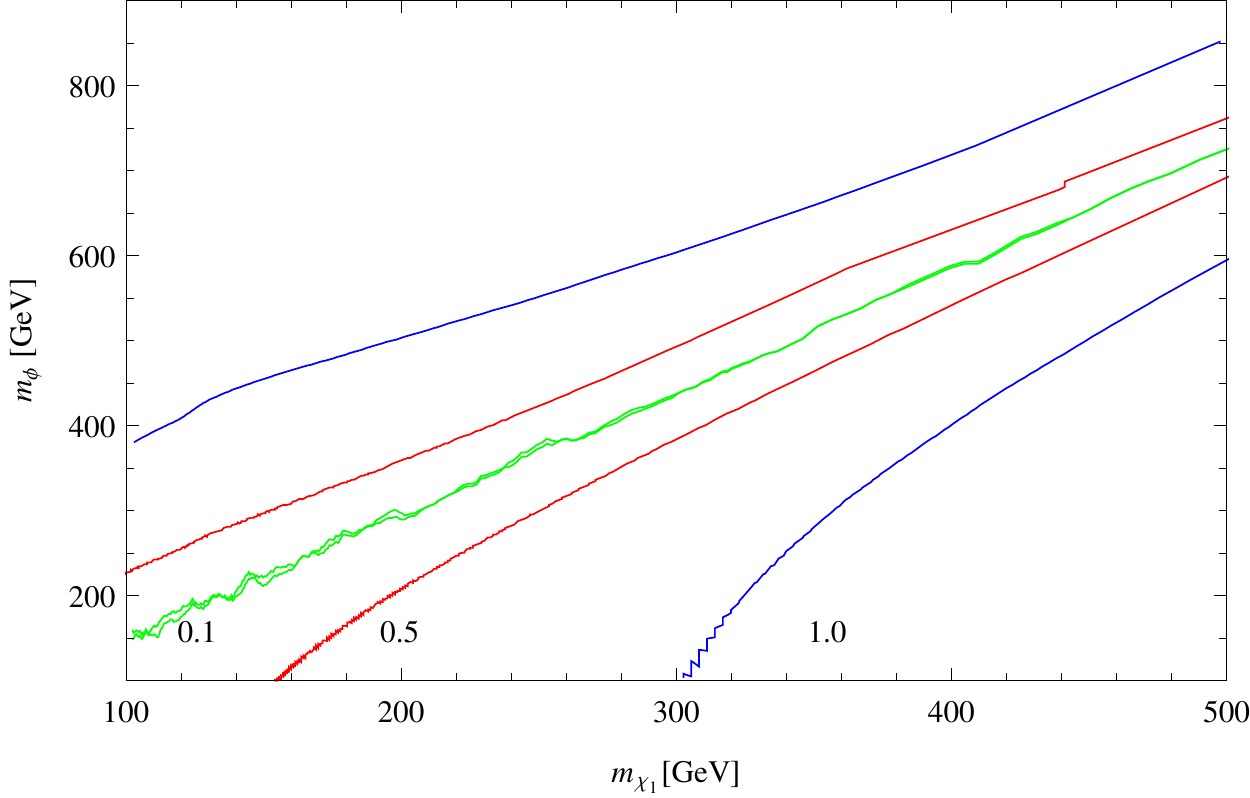}
\end{center}
\caption{Relation between familon and DM masses from the observed relic density, for the familon fields with fixed mass ratios: $m_\phi$ is the mass of $\phi_{2,3}$ and $m_{\phi_1}=\sqrt{2} m_\phi$. Fixed $u=\Lambda=1000$ GeV and $m_{\chi_{2,3}}=10\times m_{\chi_1}$. The numbers in the plot refer to $y_{\chi_i}=c_{\phi H}= 1.0,\,0.5,\,0.1$. }
\label{fig:familon-DM-mass}
\end{figure}
Now we will study the connection between DM and familon masses that is imposed from demanding to match observed relic density.
For the case of $m_{\chi_{2,3}}>1.2\,m_{\chi_1}$, a suitable combination of the familon and the DM masses, i.e.\ $m_\phi$ and $m_{\chi_1}$, respectively, always allows for the correct relic density. 
The reflection of this constraint on the mass parameter space is shown in fig.~\ref{fig:familon-DM-mass}, for the three sets of parameters: $y_{\chi_i}=c_{\phi H}= 1.0,\,0.5,\,0.1$. 
The shown lines of different color each denotes the contour line for a specific value for the coupling parameters from this set, where the relic abundance of 0.1 can be produced by adjusting the remaining parameters. For masses inside the contours,  the abundance tends to be smaller than 0.1, which can be remedied for instance by allowing for larger values of the familon VEV, smaller mass splittings between the DM fields, or a reduction of $y_{\chi_i}$ and/or $c_{\phi H}$. It is possible to identify which of the diagrams in fig.~\ref{fig:annihilationchannels} is the most relevant, i.e.\ which parameters have the most effect on the relic density, and which are the dominant SM final state particles:

On the upper branch of each contour line, the main annihilation products are pairs of Higgs scalars, and the relevant diagram is \ref{fig:annihilationchannels} {\bf b}. On the right-hand branch of the contour lines, the four-point interaction between the DM, the Higgs boson and the $\phi_1$, which corresponds to the operator in eq.~\eqref{eq:Oeff5_ell}, yields a relevant contribution to the total annihilation rate.

In fig.\ \ref{fig:familon-DM-mass}, the green line converges onto the case of resonant s-channel annihilation, which is shown by the right-hand plot in fig.~\ref{fig:overview-phi-ell}. This implies, that in order to have a DM candidate with a mass $m_\chi \sim 1$ TeV, the mass of the DM and the familon fields have to be related by the resonance condition \eqref{eq:resonancecondition}, in order not to result in overabundant DM.

\subsection{$A_4$ with neutrino-familon interactions and DM}
\label{sec:familon-nu}
\begin{table}
  \centering
  \begin{tabular}{c||cc|cc|ccc|cc}
     & $\chi$ & $\chi^c$ & $\nu^{c}$ & $L$ & $e^{c}$ & $\mu^{c}$ & $\tau^{c}$  & $H$ & $\phi_{\nu}$ \\
    \hline
    $A_{4}$ & $3$ & $1$ & $3$  & $3$  & $1$  & $1''$  & $1'$  & $1$   & $3$   \\
    $U(1)_{B-L}$ & $0$ & $2$ & $1$ & $-1$ & $1$ & $1$ & $1$ & $0$   & $-2$  \\  
    \hline
  \end{tabular}
   \caption{Field assignment within $A_4 \times U(1)_{B-L}$. New fields are gauge singlets with respect to the SM gauge group. The SM gauge charges for the SM fields are unchanged.}
  \label{tab:assignment1}
\end{table}
In this section, we consider a second familon model, that derives from the SUSY frameworks discussed in ref.~\cite{Varzielas:2012ai}.
As in the previous section, the introduced FS is $A_4$, and a global $U(1)_{B-L}$--symmetry is enforced to prevent the dark fermions, given by the $A_4$--triplet $\chi$ and the three $A_4$--singlets $\chi^c_i,\,i=1,2,3$, from mixing with the neutrinos.
The non-Abelian familon portal is given by the $A_4$-triplet SM singlet scalar $\phi_\nu$, with couplings to the right-handed neutrinos $\nu^c$, that are an $A_4$--triplet SM singlet fermion field. The familon is acting as a Majoron, and in developing a VEV, it gives a mass term to the right-handed neutrinos and breaks the lepton number.
Due to the $U(1)_{B-L}$ charge assignments in tab.\ \ref{tab:assignment1}, the candidate DM fields need to have the symmetry charge B-L($\chi + \chi^c) = +2$, in order to allow for the familon portal between SM and DM. As the DM fields must be distinguished from the RH neutrinos, we opt for B-L($\chi^c) = +2$, while B-L($\chi) = 0$. Note that these assignments prevent any invariants between the DM and $L H$, and between the DM and $\nu^c$.

\subsubsection{The scalar sector}
The SUSY framework \cite{Varzielas:2010mp, Varzielas:2012ai} whence this model is derived, requires the familon field $\phi_\nu$ to develop a VEV in the particular direction in flavour space:
\begin{align}
\langle \phi_\nu \rangle = (w,w,w)\,.
\label{eq:VEValignment-nu}
\end{align}
In this type of model the VEVs are usually obtained from the SUSY preserving superpotential terms through F-terms \cite{Varzielas:2010mp, Varzielas:2012ai}.
For the sake of simplicity, we consider the scalar potential from eq.~\eqref{scalarpotential} to apply also for the model of this section, with the substitution $\phi_\ell \to \phi_\nu$. We find, that the condition
\begin{equation}
c_2=-c_1\,,\qquad \text{and} \qquad c_{\phi_\nu H} < 0
\label{eq:scalarcondition}
\end{equation}
allows for the VEV alignment from eq.~\eqref{eq:VEValignment-nu}, with a (possible) hierarchy $w > v_\mathrm{EW}$. For instance, $c_1=c_{\phi H}=10^{-3}$ yields the magnitude of the familon VEV is $w=780$ GeV, where larger values for $w$ are possible for smaller values of $c_1,\,c_{\phi H}$.

Without considering bi-linear and tri-linear terms in the familon fields as discussed previously, mass-terms for the familons emerge after symmetry breaking with the mass matrix
\begin{equation}
M^2_{\phi_\nu} = w^2 \,
  \begin{pmatrix}
    a + c +\delta & b - c/2 & b -c/2  \\
    b - c/2 & b + c & a - c/2 +\delta \\
    b - c/2 & a - c/2 +\delta & b + c
  \end{pmatrix} \,,
\label{eq:massmatrix-nu}
\end{equation}
with the parameters $a,b,c,\delta$ being linear combinations of the VEV-magnitude $w$ and the coupling parameters $c_i,\,c_{\phi_\nu,s}$ and $c_{\phi H}$:
\begin{equation}
a = 20 c_1 +8 c_2\,, \qquad b = 8 c_1 +14 c_2\,, \qquad c = \frac{16 c_{\phi_\nu,s} }{3}\,, \qquad \delta = 2\, c_{\phi_\nu H}\,\frac{v_\mathrm{EW}^2}{w^2} \,.
\end{equation}
When the condition in eq.~\eqref{eq:scalarcondition} is relaxed to $c_2=-c_1+\varepsilon$ for $\varepsilon \ll 1$, the lightest eigenstate of the mass matrix \eqref{eq:massmatrix-nu} is given by the linear combination $\phi_{\nu_1}+\phi_{\nu_2}+\phi_{\nu_3}$ with mass given by $w\times (36 \varepsilon + \delta)^{1/2} $, which is suppressed compared to the mass scale of the other two familon fields, that is given by $w$.
In the following, we consider one physical familon field with a mass on the electroweak scale, and two degenerate familon fields with masses above the TeV scale.

\subsubsection{SM-familon interactions: Neutrinos}
The part of the Lagrangian density of the model contains Yukawa-like terms including the neutrinos, is given by
\begin{equation}
\mathscr{L}_{\nu} \supset -y_\phi \, \phi_\nu \left[ \nu^c \nu ^c\right]_s - y_H \tilde H \left[ L \nu ^c\right]_s\,,
\end{equation}
where $\tilde H = i \tau_2 H^\star$. The breaking of the $A_4$ and $SU(2)_L$ symmetries, and insertion of the respective scalar VEVs, results in Majorana and Dirac masses for the neutrino mass matrix. 

The mass eigenstates have to include the observed three light (active) neutrino fields. Further heavy (mostly sterile) neutrinos are present, that have to be either sufficiently heavy to escape experimental detection up to now or to be only slightly mixed. In the following, we consider the heavy neutrinos to have masses on the TeV scale. The right-handed neutrinos are not part of the DM; when the mixing between light (active) and heavy (sterile) neutrinos is not exactly zero, and for the here considered masses of the heavy neutrinos, $m_N$, being on or above the TeV scale, the heavy ones will decay rapidly into the light ones. 

\subsubsection{The dark sector}
The dark sector consists in a chiral fermion $\chi$, that is a SM gauge singlet and an $A_4$ triplet, and one conjugate field $\chi^c$, that is a singlet under the SM gauge group as well as under the $A_4$.
The $\chi$ fields have Yukawa-type interactions with the familon fields $\phi_\nu$
\begin{equation}
\mathscr{L}_{\phi_\nu \chi \chi} = y_{\chi} [\phi_\nu \chi] \chi^c = y_{\chi} \phi_{\nu_1} \chi_1 \chi^c + y_{\chi} \phi_{\nu_2} \chi_3 \chi^c + y_{\chi} \phi_{\nu_3} \chi_2 \chi^c\,.\,,
\end{equation}
with the ensuing masses being degenerate, once the $\phi_\nu$ acquires its VEV:
\begin{align}
  m_{\chi} =  y_\chi\,w\,.
\label{eq:chimass1}
\end{align}

\subsubsection{Dark matter relic density}
\begin{figure}
\begin{center}
\includegraphics[width=0.45\textwidth]{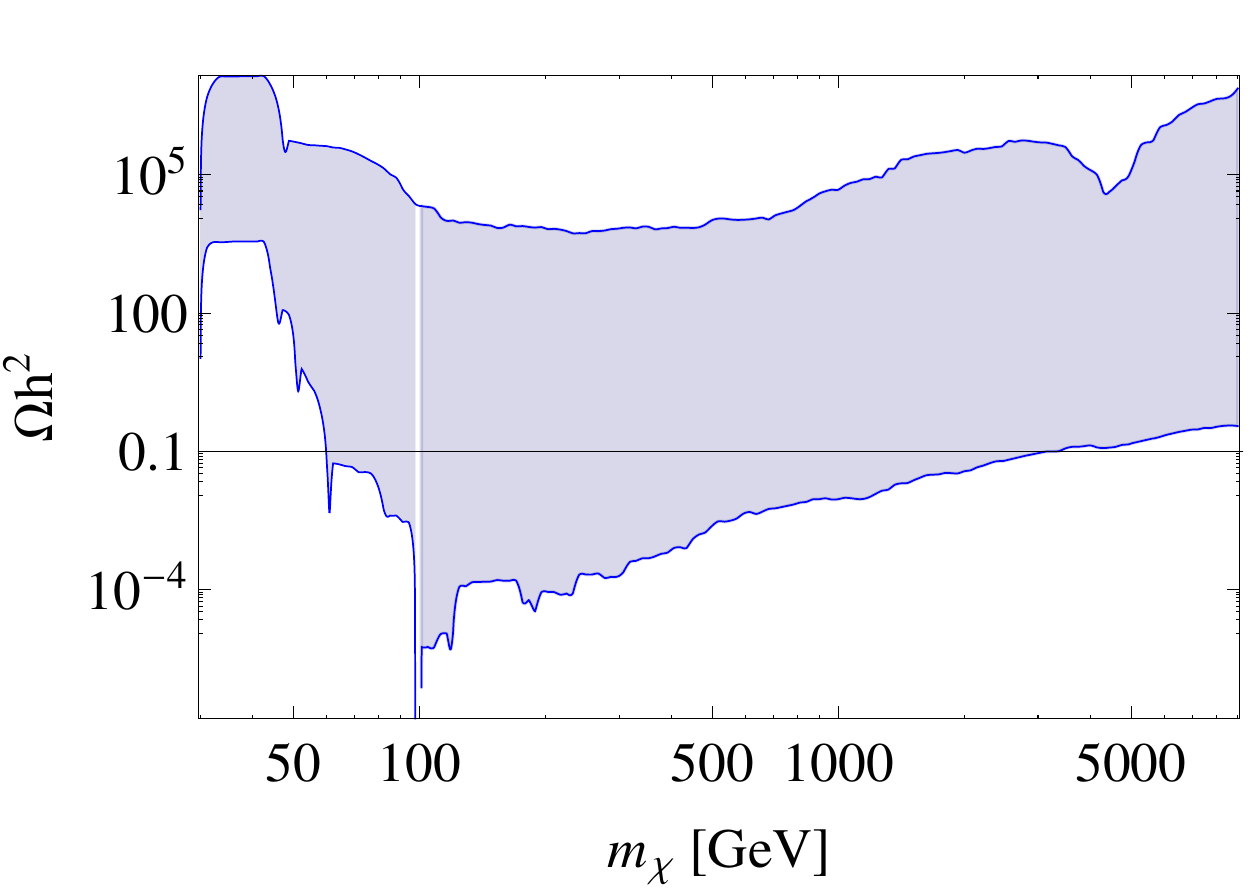}
\end{center}
\caption{Overview of the produced relic density with the non-Abelian familon portal given by $\phi_\nu$. The model parameters were marginalised within the numerical domains defined in \eqref{eq:rangenu}.}
\label{fig:DMoverview-nu}
\end{figure}

As before, we use {\tt Feynrules} to implement the model into {\tt microMEGAs} and {\tt T3PS} to perform the parameter space scans.
We confine the model parameters to the following ranges:
\begin{equation}
\begin{array}{c}
10 \text{ GeV} \leq m_\chi \leq 10 \text{ TeV} \\
100 \text{ GeV} \leq m_{\phi_\nu^{\rm light}} \leq 1 \text{ TeV} \\
0.01 \leq y_{\chi},\,y_{\phi \nu},\,c_{\phi H} \leq 2 \\ 
m_N = 100 \text{ GeV, } m_{\phi_\nu^{\rm heavy}} = 10 \text{ TeV}
\end{array}
\label{eq:rangenu}
\end{equation}
while we use for the active-sterile (or light-heavy) neutrino mixing: $|\theta|^2 = 0.001$ \cite{Antusch:2014woa,Antusch:2015mia}.
We display the resulting range for the relic density, marginalised over the remaining parameters in the ranges defined in eq.~\eqref{eq:rangenu} in fig.~\ref{fig:DMoverview-nu}. It shows, that for DM masses between 60 GeV and $\sim 4$ TeV the observation can be matched. 
In general, the abundance constraint imposes no relation between DM mass and familon mass in this model; each combination of  $m_\chi$ and $m_{\phi_\nu^{\rm light}}$ in the limits of eq.~\eqref{eq:rangenu} allows for to the correct abundance with the appropriate choice of the other parameters.

\begin{figure}
\begin{center}
\includegraphics[width=0.45\textwidth]{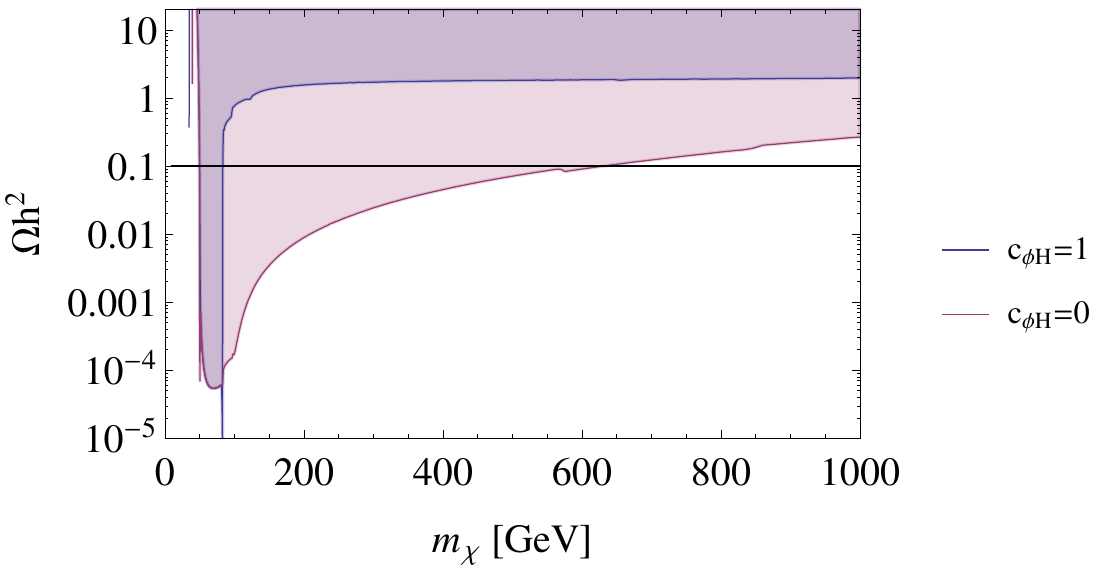}
\includegraphics[width=0.54\textwidth]{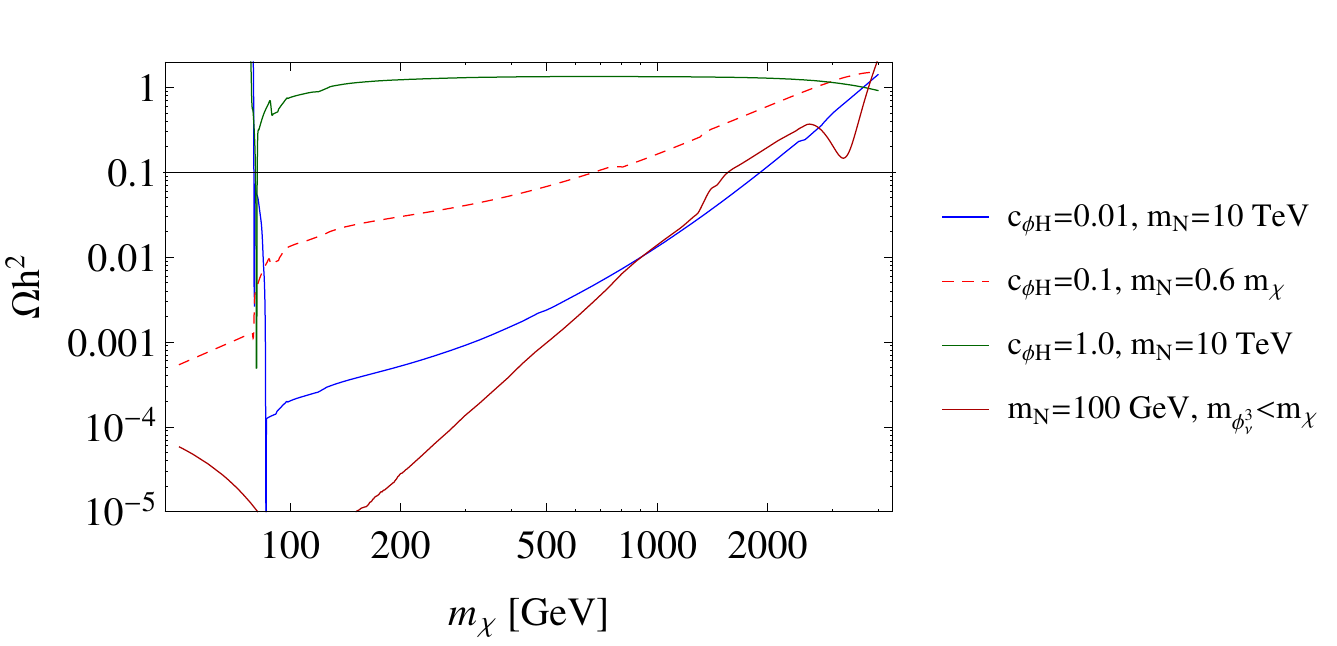}
\end{center}
\caption{{\it Left:} Red line: $c_{\phi H}=1$, Blue line: $c_{\phi H}=0.01$. All Yukawa couplings equal one, $m_N=100$ GeV.
{\it Right:} The blue line is the lower limit on the abundance. It is given by resonant s-channel production of two heavy neutrinos (being lighter than the DM). Relevant parameters controlling the cross section are the respective Yukawa couplings.
The two other colored lines are the same, just with $c_{\phi H} = 0.01,0.1$, respectively. In this case, the scalar parameters are dominating in importance over the neutrino ones.}
\label{fig:DMresannihilation}
\end{figure}
We first consider the case of $m_\chi > m_N$, and the hieararchy of the couplings $y_{\phi \nu} \gg c_{\phi H}$. In this case, the Feynman diagram that dominates the total annihilation cross section is given by fig.~\ref{fig:annihilationchannels} {\bf a}. The relic density is then minimised for $y_\chi=y_{\phi \nu}=1$ and the s-channel diagram being on resonance, i.e.\ $2\,m_\chi \sim m_{\Phi_1}$, which is shown by the red line in the left-hand plot of fig.~\ref{fig:DMresannihilation}. The red area above indicates, that the correct relic density can be obtained by tuning the couplings $y_\chi$ and $y_{\phi \nu}$, respectively.

The second relevant case to consider, is an intermediate one, where the familon couplings $c_{\phi H} = y_{\phi \nu}=1$. This case is shown --- with the heavy neutrinos masses being fixed: $m_N = 100$ GeV --- by the blue curve in the left-hand plot in fig.~\ref{fig:DMresannihilation}. We observe, that, while $m_\chi < m_N$, the minimal relic density matches that of the other case, where only annihilation into right-handed neutrinos is dominant. Passing the kinematic threshold, however, reduces the annihilation cross section compared to the red line. This comes about since the partial decay width of the familon into two heavy neutrinos, $\Gamma(\phi_\nu \to 2\nu_\mathrm{heavy})$, yields a significant contribution to the total decay width of the $\phi_\nu$.
We note that, for the above setting of parameters, the partial decay width of the familon $\phi_\nu$ into Higgs and Goldstone bosons is enhanced depending on the familon VEV and its mass, with a factor ${\cal O}(w/m_{\phi_\nu})$, which we will discuss in more detail below.

The maximal mass for the DM particles $\chi_i$, that allows the production of the observed relic density, is given when $m_N > m_\chi$ and $c_{\chi H}=1$. In this scenario, the Yukawa coupling $y_{\phi \nu}$ becomes irrelevant for the numerical evaluation. 
Althernatively, the resonant s-channel also allows for large DM masses. The possible mass range for this resonant annihilation is shown by the three curves in the right-hand plot of fig.~\ref{fig:DMresannihilation}: 
In this figure, the intermediate scenario is shown for $c_{\phi H}=0.1$ and kinematically available heavy neutrinos in the annihilation cross section, by the red line. The mass range is slightly smaller, compared with the scalar case, emphasising the power of the scalar couplings, but at the same time demonstrating, that the relic density can be matched also {\it without} strong scalar interactions, which is important to preserve the hierarchy of the VEVs, which requires small values for the scalar couplings $c_1$ and $c_{\phi H}$.
The case with the familon being lighter than the $\chi_i$ leads to the dominant contribution to the annihilation cross section being given by the t-channel process shown in fig.~\ref{fig:annihilationchannels} {\bf c}. This case is shown by the dark red line in the right plot of fig.~\ref{fig:DMresannihilation}. Notably, this process does not increase the allowed mass range for the DM.

A theoretically limiting case is given by the parameter settings that are denoted by the green line, which always stays above the observation, except for $m_\chi \lesssim M_W$.
This comes from the total decay width of the familon, $\Gamma_{\phi_\nu}$, that is enlarged by decays into two Higgs bosons, the partial decay width of which is proportional to the familon VEV $w$. The perturbative treatment of the model breaks down, when $\Gamma_{\phi_\nu} \simeq m_{\phi_\nu}$ \cite{Durand:1989zs}, such that an upper bound on the model parameters emerges:
\begin{equation}
c_{\phi H} \, w \leq m_{\phi_\nu} \sqrt{\frac{\pi}{\Pi}}\,.
\end{equation}
Here $\Pi$ is the phase space factor for the decay $\phi_\nu \to HH$.
This implies for the DM relic density, that parameter sets that are violating the above bound are meaningless in a perturbed sense, which, however, has no implication for our previous study, since they are not compatible with the abundance constraint. Equivalently one can state, that $w$ beyond the TeV scale is not very compatible with $m_{\phi_\nu}$ on the weak scale, when the familon interacts with the Higgs boson.

\subsection{GUT inspired $SU(3)$ family symmetry}
\label{sec:GUTmodel}
This model is based on an $SO(10)$ SUSY GUT broken to the Pati-Salam $SU(4) \times SU(2)_L \times SU(2)_R$ and then to the SM gauge group, combined with an $SU(3)$ FS \cite{deMedeirosVarzielas:2005ax}.
At the Pati-Salam stage the gauge group is Left-Right symmetric, and we denote the field that contains all the left-handed SM fermions as $\psi_i$ and the conjugate of the right-handed SM fields as $\psi^c_i$, where $i$ is the generation index, that is in here also the $SU(3)$ FS index. 
As the conjugate of right-handed spinors transform under the Lorentz group as left-handed spinors, Dirac mass terms can be made between $\psi$ and $\psi^c$. 

The model further contains three familon scalar fields, singlets under the gauge groups and anti-triplets under the $SU(3)$ FS, which allows to form invariants with the fermions. 
We assume that the three familons develop non-zero VEVs in different FS alignments and with different magnitudes.
For clarity, we follow the notation of \cite{deMedeirosVarzielas:2005ax} and include subscript in the name that is a reminder of the specific VEV direction for each of the familons: the $\bar \phi_3$ field has a zero in the 1st and 2nd entry $(0,0,a)$, the $\bar \phi_{23}$ field has a zero in the 1st entry and the other two with equal magnitudes $(0,b,-b)$, and the $\bar \phi_{123}$ field has no zero entries and all three entries with equal magnitudes $(c,c,c)$ \cite{deMedeirosVarzielas:2005ax}. In the following, we posit a hierarchy of the familon VEVs, $a \gg b \gg c$, which is also motivated from the GUT model.

Among the invariants that can be constructed with the field content introduced above, one can find terms that are connecting the fermions and scalars of the model, and can give rise to the masses of the charged fermions. The most relevant of these terms for our discussion are
\begin{equation}
\mathscr{L}_{\psi\phi H} \supset c_3 (\bar{\phi}_{3}^i \psi_i) (\bar{\phi}_{3}^j \psi_j^c) \frac{H}{\Lambda^2} + c_{123} (\bar{\phi}_{23}^i \psi_i) (\bar{\phi}_{123}^j \psi_j^c) \frac{H}{\Lambda^2} + c_{123} (\bar{\phi}_{123}^i \psi_i) (\bar{\phi}_{23}^j \psi_j^c) \frac{H}{\Lambda^2} \,.
\label{eq:SU3lagrangian}
\end{equation}
The equality of the coefficients of the last two terms is due to $SO(10)$, making the Yukawa structures symmetric.
Furthermore, a term exists with a Georgi-Jarlskog field $H_{45}$ that gives different Clebsch-Gordan coefficients to the charged leptons and down quarks for the 2nd generation, $(\bar{\phi}_{23}^i \psi_i) (\bar{\phi}_{23}^j \psi_j^c) \frac{H H_{45}}{\Lambda^3}$, see ref.~\cite{deMedeirosVarzielas:2005ax} for further details. We shall not consider this contribution in the following, for the sake of presentational simplicity.
After symmetry breaking, i.e.\ when the familon $\phi_3$ acquires its VEV $\langle \bar{\phi}_{3} \rangle = (0,0,a)$,
the term giving rise to the mass of the 3rd generation fermions is the most relevant one from eq.~\eqref{eq:SU3lagrangian}:
\begin{align}
\frac{ a}{\Lambda^2} \psi_3^c (\bar{\phi}_{3}^i \psi_i) \frac{(h+v)}{\sqrt{2}}\,.
\label{barphi3_int}
\end{align}
The parameter $\Lambda$ is again the mass scale of the FN messengers that were integrated out. This term allows the familon $\bar\phi_3$ to couple to the fermion current and the Higgs boson.

\subsubsection*{Familon mass structure}
As in the previously discussed examples, we first consider the case that the familon masses arise from the terms in the scalar potential that contain exclusively the familon fields and no explicit mass terms. Then we can write down the following $SU(3)$ invariant tri-linear with the Levi-Civita tensor $\epsilon_{ijk} (\bar{\phi}_{3}^i \bar{\phi}_{3}^j \bar{\phi}_{3}^k)$:
\begin{align}
  M^2_{\bar{\phi}_{3}} &= m^2
  \begin{pmatrix}
    0 & 1  & 0  \\
    -1 & 0 & 0  \\
    0  & 0  & 0
  \end{pmatrix} \,.
  \label{eq:familonmass}
\end{align}
If no other mass terms are included, the lightest familon remains exactly massless.
We use this as a motivation to consider the mass of the lightest $\phi_3$ field component to be an independent parameter, arising e.g.\ from small soft SUSY breaking mass terms such as $m^2 \bar\phi_{3}^i \bar\phi_{3_i}^\dagger$, and residing on the electroweak scale.

\subsubsection*{Dark sector}
Given that the SM fermions originate from the $SO(10)$ GUT fields, the left-handed and right-handed components have the same transformation properties under the $SU(3)$ FS. The same argument need not be true for the dark fermions: a simple option would be to have $\chi^i$ as anti-triplet and $\chi^c_i$ as a triplet, in which case they could have mass terms by themselves. This is not the case we are interested in, as we require the dark sector to interact with the familon portal. We therefore choose both the left-handed and right-handed dark fermions to transform as anti-triplets under $SU(3)$ FS, which can lead to the simple invariant term $\epsilon_{ijk} \chi^i \chi^{c j} \bar{\phi}_{3}^k$ 
and results in a mass matrix for the dark fermions of the form of eq.~\eqref{eq:familonmass}.
Explicitly, the dark fermions couple directly to the familon fields via the following Yukawa terms:
\begin{align}
\mathscr{L}_{\phi_3 \chi \chi} = - y_3 \left( \bar{\phi}_{3}^{1} (\chi^2 \chi^{c 3} - \chi^3 \chi^{c 2}) + \bar{\phi}_{3}^{2} (\chi^3 \chi^{c 1} - \chi^1 \chi^{c 3}) + \bar{\phi}_{3}^{3} (\chi^1 \chi^{c 2} - \chi^2 \chi^{c 1}) \right)\,.
\end{align}
The combination of $\chi^i \chi^{cj}$, with $i,j$ fixed, corresponds to a specific fermion mass eigenstate, which we label $\eta^{ij}$, where the superscript $ij$ is a part of the label. Thus, after symmetry breaking, two degenerate Dirac mass eigenstates $\eta^{12}$ and $\eta^{21}$ with a mass proportional to the VEV $a$ emerge, together with the massless fermion $\eta^{33}$.

Further invariants with the familons and the dark fermions can be constructed, which add to the masses of the dark sector. Keeping in mind the charges of the original GUT model that are implicit in eq.(\ref{eq:SU3lagrangian}), we consider the additional terms
\begin{align}
\mathscr{L}_{\chi\chi \phi} \supset \xi_1\, \epsilon_{ijk} \chi^i \chi^{c j} \bar{\phi}_{3}^k + \xi_2\, \frac{\bar{\phi}_{3_i}^{\dagger} \bar{\phi}_{23}^j \bar{\phi}_{123}^k}{\Lambda^2} \chi^l \chi^{c m}\,,
\end{align}
which contribute to the fermion masses after symmetry breaking, although their contribution is suppressed by the VEV hierarchy and the FN messenger scale, from the UV-complete theory. The presence of the second term is necessarily allowed by the symmetries of the original model, given that the combination of $\bar{\phi}_{3} \bar{\phi}_{3}$ has the same overall charge as $\bar{\phi}_{23} \bar{\phi}_{123}$ (see eq. (\ref{eq:SU3lagrangian})).
The $\xi_i$ above are ${\cal O}(1)$ coefficients, and the indices $i, j, k, l, m$ have to be contracted in all possible ways ($i$ with one of the others, and the remaining contract with one $\epsilon$ tensor).
Including these contributions, it is easy to find parameter settings such that the masslessness of $\eta^{33}$ is lifted and a mass of ${\cal O}(v_\mathrm{EW})$ can be realised.

It is important to notice, that the contributions from the VEVs $\langle\bar{\phi}_{23}\rangle$ and $\langle\bar{\phi}_{123}\rangle$ to the mass matrix render the mass eigenstate $\eta^{33}$ a linear combination of the flavour eigenstates $\chi^i$ and $\chi^{cj}$:
\begin{equation}
\eta^{33} = s_{\theta_1} \chi^1 \chi^{c2} + s_{\theta_2} \chi^2 \chi^{c1} + c_\theta\, \chi^3 \chi^{c3}\,,
\label{eq:mixing}
\end{equation}
with the coefficients $s_{\theta_{1,2}}\ll1$, such that we approximate $\theta_1 = \theta_2 = \theta/2$, with the mixing angle $\theta$ being a function of the VEV-ratios that mixes the DM fields.
Since light familon field $\bar{\phi}_3^{3}$ couples to $\chi^1 \chi^{c2}$ and $\chi^2 \chi^{c1}$, with the mixing in eq.~\eqref{eq:mixing}, the physical DM field $\eta^{33}$ also interacts with the light familon field, with a coupling strength proportional to $s_\theta\, y_3$.

\begin{figure}
\begin{center}
\includegraphics[width=0.45\textwidth]{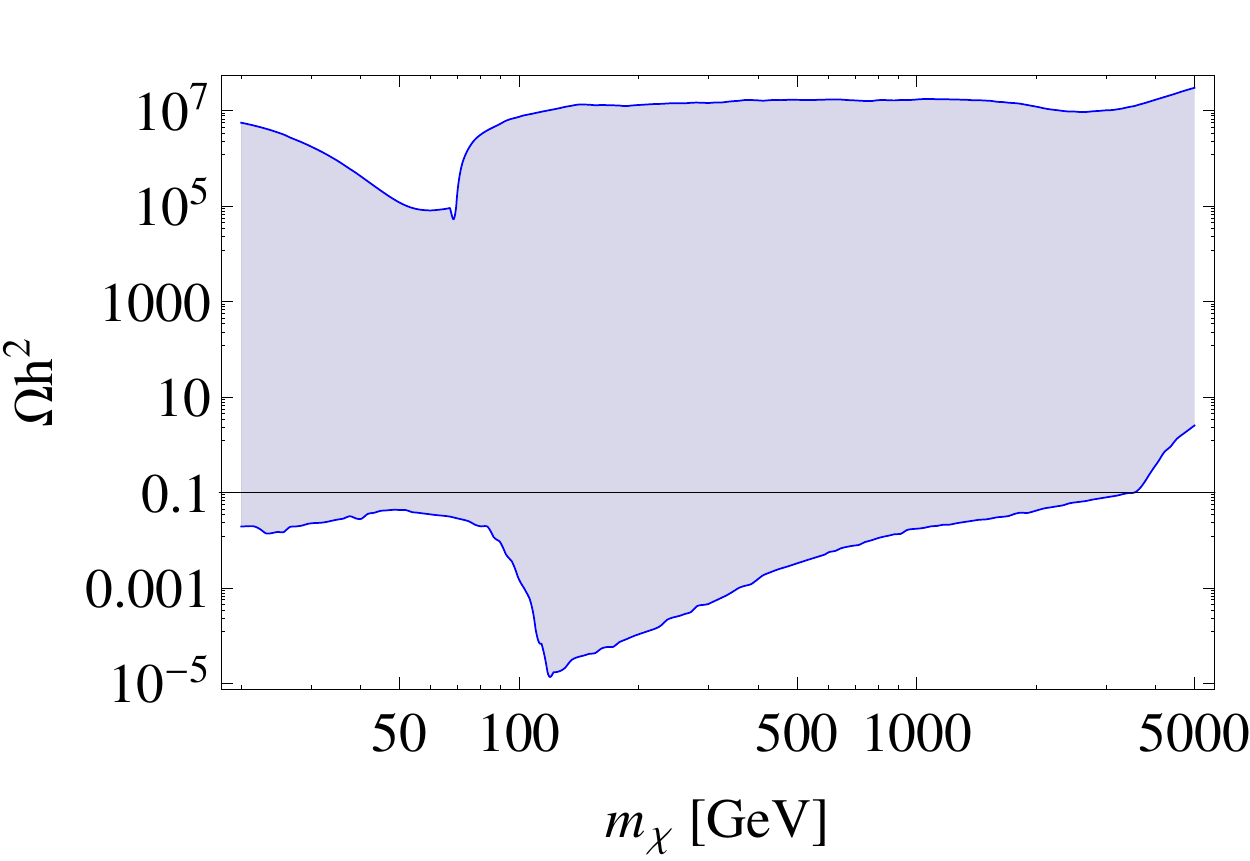}
\includegraphics[width=0.45\textwidth]{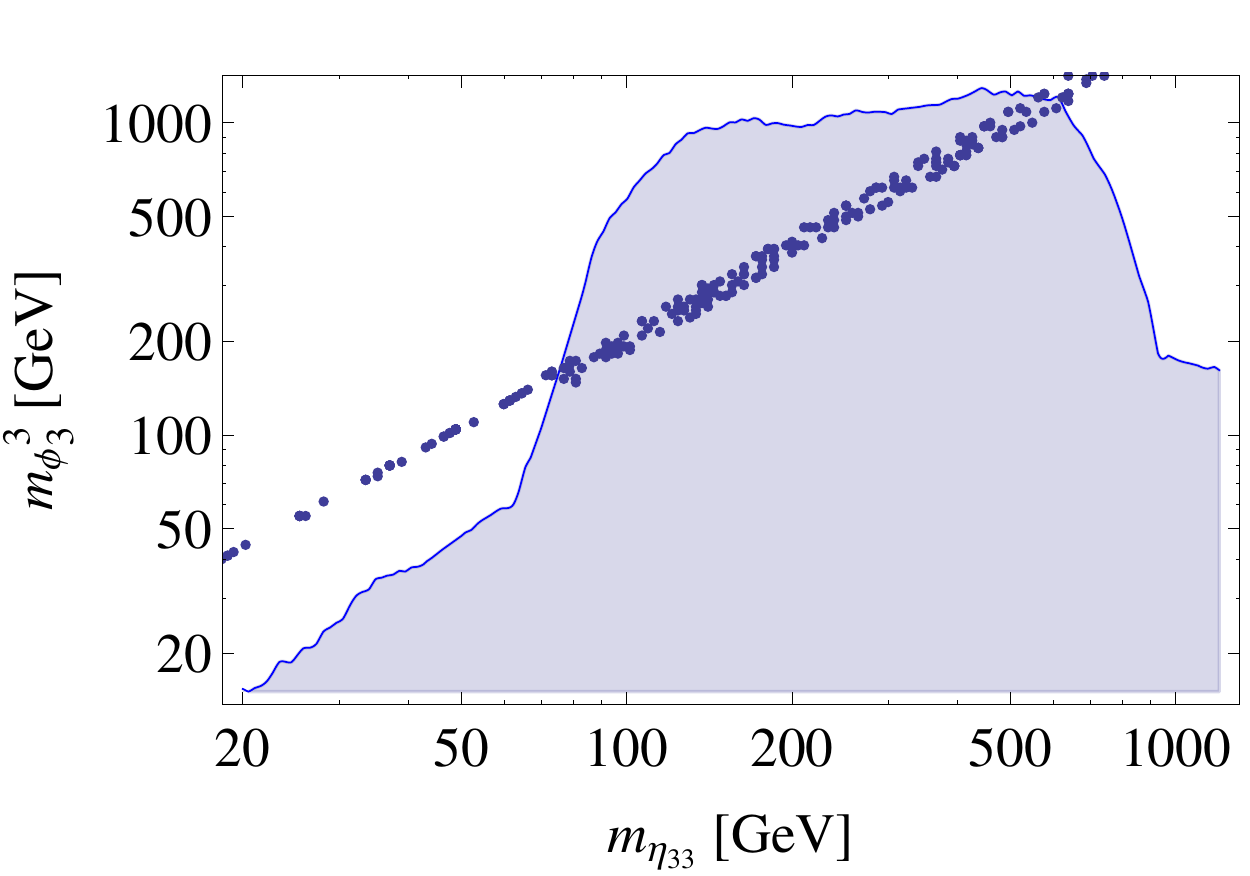}
\end{center}
\caption{{\it Left:} Overview of the produced relic density with the non-Abelian familon portal given by the third component of $\bar{\phi}_3$, with the model parameters in the domain defined in eq.~\eqref{eq:domainGUT}. The black horizontal line denotes the value of the observed relic density. {\it Right:} Relation between the DM candidate mass and the lightest familon from the requirement that the observed relic abundance is matched. The blue dots denote the fine-tuned scenario with resonantly annihilating DM, which implies $m_\phi\sim 2 m_\chi$.}
\label{fig:familonGUT0}
\end{figure}
The relic density produced in a scatter scan with {\tt T3PS} and {\tt MicrOMEGAs} is shown in the left panel of fig.~\ref{fig:familonGUT0}. Therein, the relevant model parameters have been varied in the following ranges:
\begin{equation}
\begin{array}{c}
10 \text{ GeV} \leq m_{\eta^{33}},\,m_{\phi_3^{3}} \leq 10 \text{ TeV,} \\
0.001 \leq s_\theta\, y_3 \leq 0.1\,, \\
0.01 \leq c_{\phi_3 H} \leq 1\,,
\end{array}
\label{eq:domainGUT}
\end{equation}
while $\Lambda=a=10$ TeV and the other DM particles $\eta^{12},\,\eta^{21}$ as well as the heavy familon components $\bar\phi_3^{1,2}$ have been neglected, considering their masses to be above $\sim 10$ TeV, for simplicity.
The right panel of fig.~\ref{fig:familonGUT0} shows the masses of the familon $\bar\phi_3^{3}$ and the DM $\eta^{33}$ that are compatible with the abundance constraint.
Larger values for the masses are possible, when the resonant s-channel annihilation is considered, which is shown by the blue dots in the same figure. In this case, masses for DM and familon up to 10 TeV are possible, but the simplifications due to the masslessness of $\bar\phi_3^{3}$ and $\eta^{33}$ lose their validity.

\section{Collider phenomenology of non-Abelian Familon fields}
\label{sec:colliderpheno}
We discuss implications and signals at high energy colliders in this section, for more complete investigations, see \cite{Tsumura:2009yf}. The recent and thorough collider analysis in ref.\ \cite{Berger:2014gga} considers a low energy theory from a specific UV-complete flavour model, with particular emphasis on the mixing between the (lightest) familon and the Higgs boson. 
Here, we consider the UV-cutoff to be above the TeV scale, and the resulting hierarchy of the electroweak and FS VEVs to suppress the scalar mixing between the Higgs boson and the familons, such that other observables become relevant. In the following, we will investigate some of those observables, with a focus on high-energy colliders.

\subsection{Effective framework for non-Abelian flavour symmetries}
An effective interaction term between the quark current and the lightest familon field, that can be generated after electroweak symmetry breaking, has the form
\begin{equation}
\mathscr{L}_{\phi qq} \supset f_{q q^\prime} \phi \, \bar q\, q^\prime\,,
\label{eq:eff_familonquarkcoupling}
\end{equation}
with the familon $\phi$ and the quarks $q$ and $q^\prime$ not necessarily of the same flavour, and $f_{q q^\prime}$ being an effective coupling constant.

The above effective quark-familon interactions take place for instance in the GUT inspired model, see eq.~\eqref{eq:SU3lagrangian}. They can also arise from adding quark-familon interactions to the model from section \ref{sec:familon-nu}, for instance when the quarks are embedded into $A_4$-triplets like the leptons:
\begin{equation}
\mathscr{L}_{\phi q} \supset f_{\phi q} \left(c_1[\phi q]_1 [\phi q]_1 + c_2 [\phi q]_{1^\prime}[\phi q]_{1^{\prime\prime}} + c_3[\phi q]_{s}[\phi q]_{s}+c_4 [\phi q]_{a}[\phi q]_{a} + c_5[\phi q]_{s}[\phi q]_{a} \right)\,.
\label{eq:familonquarkcoupling}
\end{equation}
We have introduced the effective ``messenger coupling'' $f_{\phi q} = v_\mathrm{EW}/\Lambda$, with $\Lambda$ absorbing the coupling and mass of the messenger field from the UV complete model that has been integrated out.
After symmetry breaking the familon develops a VEV, which we will denote by $w$ in the following and assume to reside above the TeV scale.
The VEV hieararchy results in the quark-familon interactions to be suppressed by ${\cal O}(v_\mathrm{EW}/w)$.

As shown in section \ref{sec:DMstudy}, in models with a non-Abelian FS it is possible to have light familon fields (with masses on the electroweak scale), together with cutoff scales above the TeV scale and a successful DM candidate.
A promising discovery channel for very light familon masses, meaning $m_\phi < m_t$, and flavour violating familon-quark couplings, is given by the decay of the top quark to a charm and a light familon, which has already been investigated for instance in ref.~\cite{Babu:1999me}.

\subsection{Production mechanisms}
\begin{figure}
\begin{center}
\includegraphics[width=0.7\textwidth]{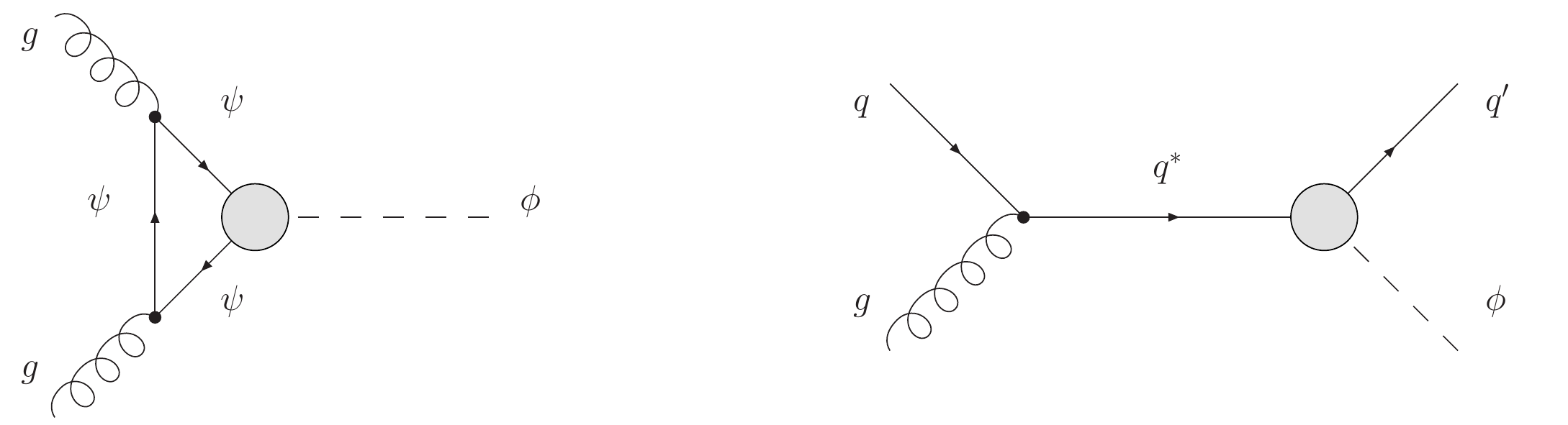}
\end{center}
\caption{The production mechanisms. {\it Left}: Gluon fusion. {\it Right:} Familon strahlung.}
\label{fig:productionmechanisms}
\end{figure}

The different processes from which a (light) familon can be produced are shown in fig.~\ref{fig:productionmechanisms}, where the gray blobs denote the effective vertices between the familon and the fermions, which are generated when the FN messenger from the UV-complete theory is integrated out.
In the framework of eq. (\ref{eq:familonquarkcoupling}), the effective vertex from eq. (\ref{eq:eff_familonquarkcoupling}) is given by:
\begin{equation}
f_{qq^\prime} = c_{qq^\prime}\, f_{\phi q}\,w\,,
\label{eq:effectivecoupling}
\end{equation}
where $c_{qq^\prime}$ denotes the coupling constant(s) between the familon and the quarks under consideration, given by the sum over the relevant $c_i$ from eq.~\eqref{eq:familonquarkcoupling}, and $w$ is the VEV of the familon field.

The effective interaction of familons with the quark currents allows the production of the familon fields via gluon fusion, analogously to the Higgs boson. The resulting expression is proportional to the corresponding one in the SM, with  a proportionality factor of $\sum_{q\,q^\prime}f_{qq^\prime}^2$. 
Therefore, provided that $\sum_{q\,q^\prime} f_{qq^\prime}={\cal O}(1)$, the production cross section for familons via gluon fusion at $\sqrt{s} = 7$ TeV can be as large as ${\cal O}(10$ pb) for $m_\phi = 150$ GeV, while it drops down to ${\cal O}(0.01$ pb) for $m_\phi = 1$ TeV.

\begin{figure}
\begin{center}
\includegraphics[width=0.45\textwidth]{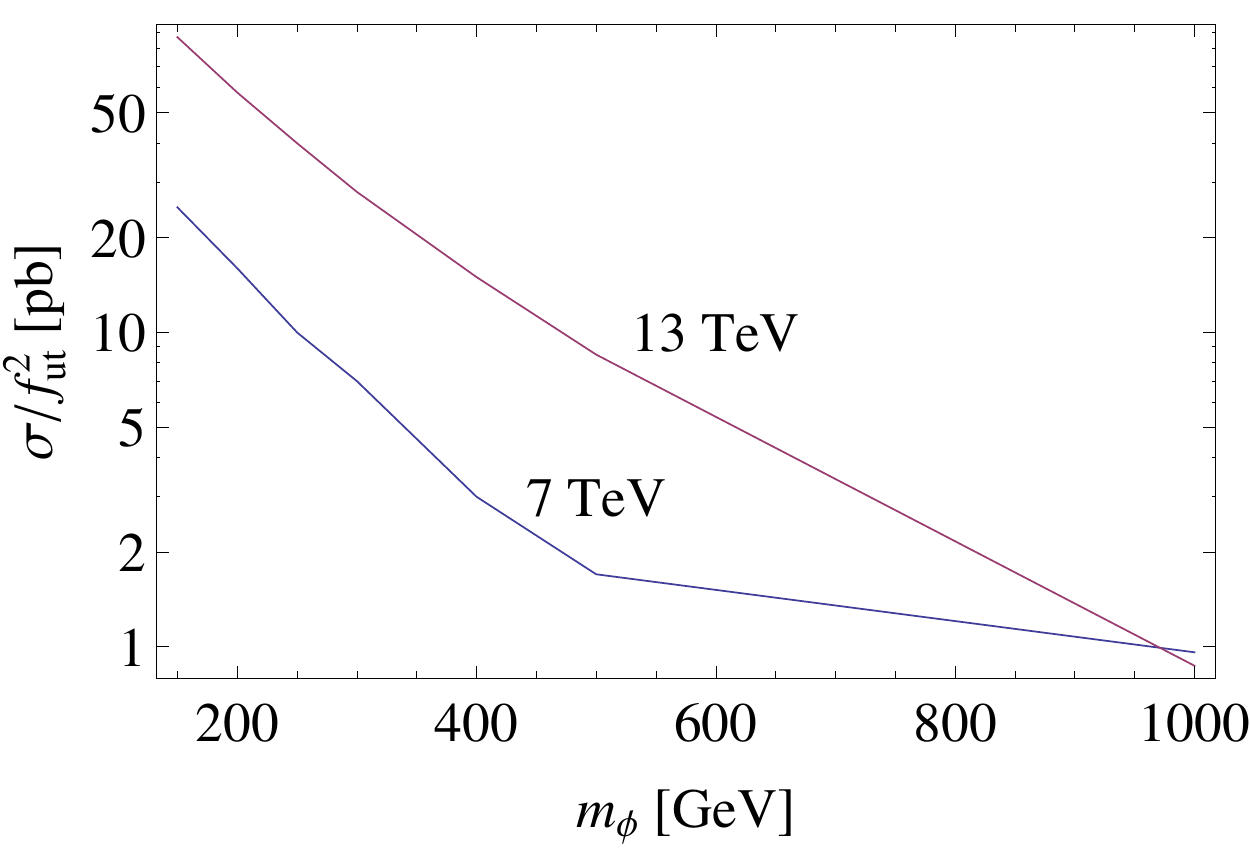}
\end{center}
\caption{Production cross section for familon strahlung at the LHC from partonic initial states, normalised by the effective familon-quark coupling. In this figure, the CTEQ parton density functions have been used.}
\label{fig:familonstrahlung}
\end{figure}
Another production mechanism for familons is familon strahlung, which occurs when a quark changes flavour while radiating off a familon, in particular when the initial quark's mass is larger than the familon mass \cite{Tsumura:2009yf}.
Familon strahlung can be the most efficient production mechanism for the familons. From the familon-quark interaction as defined in eq.~\eqref{eq:familonquarkcoupling}, we find that the dominant contribution to familon production stems from the $\phi u t$ coupling.

The production cross section for familon strahlung, normalised by $f_{u t}^2$, is shown in fig.~\ref{fig:familonstrahlung}. We note, that for $\sqrt{s} = 7$ TeV and 13 TeV, with $f_{u t} =1$ and $m_\phi=150$ GeV, a production cross section of $25 \times f_{ut}^2$ pb and $87\times f_{ut}^2$ pb, respectively, are possible.

There is one more contribution to familon production, that comes from the mixing term in the scalar potential, $H^\dagger H\,[\phi \phi^\dagger]$.
It appears pointless, however, to search for familons produced from Higgs bosons, because the latter have themselves rather small production cross sections. It turns out, that the production of familons in the fusion of two Goldstone bosons, i.e.\ the longitudinal modes of the weak gauge bosons, are completely negligible compared to the two aforementioned production mechanisms. This is as expected, since the Goldstones are coupling to the parton current $q$, which results in a factor $m_q^2/s$, with $s$ being the square of the center of mass energy of the parton, and $m_q$ its mass.

\subsection{Higgs production and decay}
The interactions  between the Higgs boson and the familon allow the heavier among the two fields to decay into the lighter one.
When the mass of the lightest familon $m_\phi$ is less than half the Higgs mass, the partial decay width of the Higgs boson into $2 \, \phi$, $\Gamma_{h\to 2\phi}$, adds to the Higgs total decay width.
The determination of the Higgs total decay width at the LHC results in $\Gamma_{\rm tot} = 6.1^{ +7.7}_{-2.9}$ MeV \cite{Barger:2012hv}. Compared to the SM prediction of $\Gamma_{\rm tot,\,SM} = 4.07$ MeV, this yields the upper bound $\Gamma_{h\to 2\phi}< $9.7 MeV, which in turn limits the product of the familon Higgs coupling and the familon VEV:
\begin{equation}
c_{\phi H}\, w < 100 \text{ GeV}\,.
\end{equation}
This constraint on the familon-Higgs coupling together with the constraints from the DM abundance observation in section \ref{sec:familon-nu} imply, that, for $m_\phi < m_H/2$, the observed relic density can only be accounted for when the DM is heavier than the sterile neutrinos and the familon-neutrino interactions are very strong.

The complementary situation, given by $m_\phi > 2 m_H$ allows the familon to decay into two Higgs bosons, which adds to the number of di-Higgs events, which provide the important measurement of the triple-Higgs coupling at the LHC and all future colliders. 
In order to estimate the number of familon-produced di-Higgs events, $N_{2H}^\phi$, we use the production cross section $\sigma_\phi(m_\phi)$ at $\sqrt{s} = 7$ TeV from fig.~\ref{fig:familonstrahlung} and the integrated luminosity ${\cal L}_{int} \simeq 20$ fb$^{-1}$, which yields 
\begin{equation}
N_{2H}^\phi = \text{Br}(\phi \to H\,H)\,  f_{qq^\prime}^2 \times
\begin{cases} 
 {\cal O}(10^6) \text{ for } m_\phi = 150 \text{ GeV} \\
 {\cal O}(10^4) \text{ for } m_\phi = 1 \text{ TeV,}
\end{cases}
\end{equation}
for two exemplary values of the familon mass, and with Br$(\phi\to H\,H)$ being the branching ratio of the familon into two Higgs bosons, one of which may be virtual, if $m_\phi < 2\,m_H$.

The SM cross section for di-Higgs production (at next-to-next-to leading order) is about 10 fb at 8 TeV center-of-mass energy, and about 40 fb at 14 TeV \cite{deFlorian:2015moa}.
The non-observation of discrepancies in the SM predicted number of di-Higgs events in run I at the LHC, $N_{2H}^{\rm SM}\sim 200$, requires that $N_{2H}^\phi$ must be small compared to $N_{2H}^{\rm SM}$. In order to quantify what we mean by ``small'', we impose the condition $N_{2H}^\phi \leq \sqrt{N_{2H}^{\rm SM}}$, which means that the resulting number of familon-produced di-Higgs events has to be smaller than the statistical one sigma fluctuation of the SM-predicted number of events, which translates into the bound 
\begin{equation}
f_{qq^\prime} \text{Br}(\phi \to H\,H) \leq  
\begin{cases} 
 {\cal O}(10^{-5}) \text{ for } m_\phi = 150 \text{ GeV} \\
 {\cal O}(10^{-3}) \text{ for } m_\phi = 1 \text{ TeV,}
\end{cases}
\label{eq:higgs-constraints}
\end{equation}
which can be used to constrain the parameters of the UV complete model. For instance eq.~\eqref{eq:effectivecoupling} implies that, if we assume all coefficients to be equal to one, we get with $\Lambda = w \geq {\cal O}(100)$ TeV for $m_{\phi}=150$ GeV, and $w \geq {\cal O}(10)$ TeV for $m_{\phi}=1$ TeV.

\subsection{Lepton flavour violation and lepton non-universality}
The violation of lepton flavour (LFV) in the final states at high-energy colliders provide another very promising signal for non-Abelian FS and the associated familon fields. For instance the model in section \ref{sec:GUTmodel} would lead to flavour violating final state fermions and an associated Higgs boson in the final state, which does not have a SM background.
At lepton colliders such as the planned electron-positron mode of the Future Circular Collider (FCC-ee) \cite{Gomez-Ceballos:2013zzn}, the Circular Electron Positron Collider (CEPC) \cite{Ruan:2014xxa} or the International Linear Collider (ILC) \cite{Baer:2013cma}, this final state could be studied with great precision, for instance when investigating the Higgs boson properties at center-of-mass energies of $\sim 240$ GeV. 
We can approximate the LFV violating cross section with
\begin{equation}
\sigma_{\phi,LVF} = \frac{\left(f_{e \mu} f_{e \tau}\right)^2\,s}{16\,\pi\,m_\phi^4}\,,
\end{equation}
with the effective low energy LFV electron-lepton couplings $f_{e \ell}$ for $\ell = \mu,\tau$. Assuming no background and a perfect detector environment, we demand ${\cal O}(10)$ LFV events for a discovery. At the FCC-ee, with a total integrated luminosity of 10 ab$^{-1}$ the discovery is feasible when both $f_{e\mu}$ and $f_{e\tau}$ are order one, and $m_{\phi} \leq 3$ TeV. If both effective couplings are of the same magnitude as the $\tau$ Yukawa coupling, i.e.\ ${\cal O}(0.01)$, a discovery of LFV would still be possible for $m_{\phi}\leq 300$ GeV. As soon as at least one of the effective familon couplings is as small as the muon Yukawa coupling or below, the familon would have to be lighter than 10 GeV in order to produce the required number of signal events.

Non-Abelian FSs often give rise to LFV decays of the Higgs boson at the loop level, the branchings for which are expected to be smaller than $10^{-2}$ \cite{Varzielas:2015iva}. Since also these decays are free of SM background, and the FCC-ee (and CEPC) are estimated to produce ${\cal O}(10^6)$ Higgs-boson events, a detection of those decays might indeed be possible.

Another observable that is sensitive to LFV is given by the charged lepton flavour violating (cLFV) decays, to which non-Abelian FSs contribute at the one-loop level.
As discussed in ref.~\cite{Varzielas:2010mp}, a very relevant constraint comes from the experimental non-observation of the process $\mu \to e \gamma$, that receives contributions from loops of familon and FN messengers, whenever the familon couples off-diagonally to charged leptons. The contribution from the non-Abelian FS to the cLHV decay rate yields an estimated lower bound on the mass scale of the FN messengers  \cite{Varzielas:2010mp} 
\begin{equation}
\Lambda \geq {\cal O}(10^4)  \text{ GeV}
\end{equation}
which does not change much with the recent update on $Br(\mu\rightarrow e\gamma)$ of $5.7 \times 10^{-13}$ from the MEG collaboration \cite{Adam:2013mnn}.

Furthermore, the measurement of lepton universality provides a generic means of assessing new physics couplings to the lepton sector. It amounts to measuring the weak interactions of the leptons in the decay modes of various mesons, but also of the $\tau$ and $\mu$ decays themselves. 
The here considered non-Abelian FS models by themselves do not lead to a sizeable modification of lepton universality observables, due to the familon-lepton couplings being on the order of the Yukawa couplings, and the interactions being further suppressed by the messenger scale.
However, as was shown in ref.~\cite{Varzielas:2015iva}, the introduction of leptoquarks to FS models makes it possible to explain the recent hints for lepton non-universal flavour decays from the LHCb measurement  \cite{Aaij:2014ora} 
\begin{align} \label{eq:RKdata}
R_{K}^{\rm LHCb} =0.745 \pm^{0.090}_{0.074} \pm 0.036
\end{align}
where $R_K \equiv \mathcal{B}\left( B \to K \mu \mu \right) / \mathcal{B}\left( B \to K e e \right)$.
The FS and familon VEVs considered here very naturally justify the textures of the leptoquark Yukawa couplings that account for the lepton flavour isolation textures proposed in \cite{Hiller:2014yaa}, or other viable data-driven patterns for the leptoquark Yukawa couplings. It is interesting that the theory-driven, FS-inspired patterns match so well the data-driven patterns imposed on leptoquark Yukawa couplings by $R_K$, LFV bounds, and rare quark decays \cite{Varzielas:2015iva}.

\subsection{Familon strahlung and single top production}
From the GUT inspired model in section \ref{sec:GUTmodel} and eq.(\ref{barphi3_int}), the following interaction terms between the familon fields and the quark currents are obtained
\begin{equation}
\begin{array}{cc}
\bar{\phi}_3^{1}: & (q_1) (q_3^c) + (q_3) (q_1^c) \\
\bar{\phi}_3^{2}: & (q_2) (q_3^c) + (q_3) (q_2^c) \\
\bar{\phi}_3^{3}: & 2 (q_3) (q_3^c) \,,
\end{array}
\label{eq:couplingterms}
\end{equation}

As discussed in section \ref{sec:GUTmodel}, the breaking of the FS, eq.~\eqref{eq:couplingterms}  gives rise to a light familon (see eq.(\ref{eq:familonmass})) that consists mostly in $\bar{\phi}_3^{3}$, and two heavy familons that consist dominantly of the fields $\bar{\phi}_3^{1}$ and $\bar{\phi}_3^{2}$. The light familon eigenstate will therefore couple mostly to $(q_3) (q_3^c)$, with a small coupling to $(q_1) (q_3^c) + (q_3) (q_1^c)$.
Moreover, in the quark mass basis we find the heaviest quark mass eigenstate to consist dominantly of $q_3 q_3^c$, with small fractions of $q_{1,2}$ and $q_{1,2}^c$.

These mixing effects imply that the off-diagonal interaction of the lightest familon mass eigenstate with the quark mass eigenstates has to be $\ll1$.
In order to assess the effective familon-up-top coupling constant $f_{ut}$, we assume that the masses of the lighter quark eigenstates and the quark mixings stem from the $b,c$ VEVs of the other familons \cite{deMedeirosVarzielas:2005ax}, which leads to the rough estimate of $f_{ut} \sim bc/a^2 \sim 0.01$. We note, that this still allows for the sizeable cross-section for familon strahlung of order femtobarn, see fig.\ \ref{fig:familonstrahlung}, which is an order-of-magnitude prediction rather than an upper bound, due to the assumptions made. 

The constraints from the DM analysis shown in fig.\ \ref{fig:familonGUT0} suggest that $m_{\phi}<m_\chi$, where $\chi$ represent the potential family of dark fermions, which implies that the decays of the light familon lead to the detectable final states being SM fermions of the third family, (and Higgs bosons,) if kinematically allowed, a fraction of which can be flavour violating.

The familon-quark interactions therefore give rise to an excellent search channel at the LHC, namely the search for a resonance in the invariant mass of the decay products of the familon, for instance in a di-jet, that is associated with a single top.

We try to estimate the prospects of detecting such a resonance by comparing event counts.
During run I at the LHC, up to ${\cal O}(10^4)$ or ${\cal O}(10^3)$ signal events might have been produced, that consist in a single top plus a familon with $m_\phi = 150$ GeV and 1.0 TeV, respectively.

In comparison, the measured single top production cross section of $\sim 115 \pm 2$ pb (CMS \cite{Chatrchyan:2012ep, Chatrchyan:2012zca} and ATLAS \cite{Aad:2012ux, Aad:2012xca, Aad:2015upn}), which is compatible with the SM prediction \cite{Kroninger:2015oma}, yields a total event sample of $n_{t}\simeq 2\times 10^6$ events. 
Since the statistical fluctuations of this number are of the same order of magnitude as the number of hypothetical signal events, it might be possible --- with a systematic analysis of the kinematic distributions --- to extract the resonance already from the present data.
Since the production cross sections increase at 13 TeV, the chances of finding light familons are enhanced at run II of the LHC, and also by its luminosity upgrade.

\section{Conclusions}
\label{sec:conclusions}
Family symmetries are well motivated through the possibility of explaining the observed patterns of fermionic masses and mixing, the so-called flavour problem of the Standard Model, and they can easily provide a connection to viable dark sectors in order generate the observed dark matter relic density through the familon portal.

In this paper we have studied three examples from the class of models with non-Abelian family symmetries, with the dark sector consisting in fermions that are gauge singlets under the Standard Model gauge group, and transform non-trivially under the family symmetry, which gives rise to (non-Abelian) familon portal interactions between the visible and the dark sector.

We have shown that regions in the parameter space of those models exist that allow for matching the observed dark matter relic density, even without the fine-tuning of resonant annihilation, and for dark matter and familon masses on the experimentally accessible electroweak scale, with the UV cutoff above $\sim$ 10 TeV.
In particular, the grand unification inspired model allows for masses of familon fields and dark matter candidate as small as ${\cal O}(10)$ GeV, the model with familon-charged-lepton interactions requires dark matter masses above 100 GeV and the model with familon-neutrino interactions allows for masses of $\sim50$ GeV, such that the observed abundance can be matched. On the other hand, dark matter masses beyond a few TeV lead to overabundant dark matter for the models considered here, unless the fine-tuned case of a resonant annihilation cross section is considered.

Concerning collider phenomenology at the LHC, we have investigated interesting signatures, such as flavour violation in fermionic final states, additional di-Higgs events and single tops, which allow to directly test the effective coupling between the familon fields and the quarks. In the case of the familon fields coupling exclusively to the lepton sector, it is possible to test lepton flavour violating leptonic final states at future lepton colliders, along with differences in the tri-linear Higgs coupling or its decay width. It is also possibile to test the low-energy effects from the UV complete models with a non-Abelian family symmetry, such as charged lepton flavour violating decays and the violation of lepton universality in meson decays.

An intriguing possibility is given by the combination of the constraints on the mass and coupling parameters from section \ref{sec:DMstudy}, which amounts to a light familon field with potentially flavour violating couplings to the fermions, with the results from collider phenomenology from section \ref{sec:colliderpheno}, which enhances the predictivity of the respective model.
We have shown that the potential of causing flavour violating signals at the LHC or at a future lepton collider is compatible with the constraints on the dark sector, and it is already subject to constraints from the collider experiments and precision experiments such as MEG.

In particular we find that, on the one hand, the bounds from the colliders effectively push the familon vacuum expectation values of the effective models to higher values, in order to accomodate the non-observation of the considered familon-induced excesses in collider data. On the other hand, the fact that the effective parameters are subject to constraints tells us, that we can expect to start seeing excesses that might hint at familons, provided the effective couplings are order one and the flavour-symmetry-breaking scale is not too far above the electroweak scale.

Since the family symmetries are introduced to explain the observed pattern of masses and mixings of the Standard Model fermions, they tend to lead to similar predictions with respect to flavour observables. It would be interesting to see, whether the phenomenology investigated here allows to distinguish between the two classes of family symmetry given by the Abelian and the non-Abelian case. The prospects for future investigations of model specific predictions from family symmetries for collider phenomenology are very encouraging.

\section*{Acknowledgements}
This project has received funding from the Swiss National Science Foundation.
This project has received funding from the European Union's Seventh Framework Programme for research, technological development and demonstration under grant agreement no PIEF-GA-2012-327195 SIFT.

\appendix
\section{$A_4$ rules \label{app:A4}}
$A_4$ has 4 irreducible representations, 1 triplet and three singlets. $1$ is the trival singlet, and $1'$ and $1''$ (non-trivial) are conjugate to one another, transforming under a specific $A_4$ generator, $T$, by getting multiplied respectively by $\omega^2$ and $\omega$ ($\omega\equiv e^{i 2 \pi/3}$, with $\omega^3 \equiv 1$).
The products are $(1 \times 1')$, $(1 \times 1'')$, $(1' \times 1'')$ transforming as $1'$, $1''$, and $1$ respectively.
The group acts on triplets through $3\times 3$ matrices. For specific triplets $A=(a_1,a_2,a_3)$, $B=(b_1,b_2,b_3)$, under generator $T$, (diagonal in the basis we are considering), $T A = (a_1, \omega^2 a_2, \omega a_3)$ (same for $B$).
The conventions follow \cite{Varzielas:2010mp, Varzielas:2012ai} with square brackets indicating $A_{4}$ products:
\begin{align}
\left[ A B \right]= &(a_1 b_1 + a_2 b_3 + a_3 b_2) \sim 1 \, , \\
\left[ A B \right]' = &(a_1 b_2 + a_2 b_1 + a_3 b_3) \sim 1'  \, ,\\
\left[ A B \right]'' = &(a_1 b_3 + a_2 b_2 + a_3 b_1) \sim 1'' \,.
\end{align}
It is also possible to construct a symmetric ($s$) and an anti-symmetric ($a$) triplet:
\begin{equation}
        [AB]_s =\frac{1}{3}\left(\begin{array}{c}
                                     2a_1 b_1-a_2 b_3-a_3 b_2\\
                                     2a_3 b_3-a_1 b_2-a_2 b_1\\
                                     2a_2 b_2-a_3 b_1-a_1 b_3\\
                                     \end{array}
                               \right) \, , \quad
        [AB]_a=\frac{1}{2}\left(\begin{array}{c}
                                 a_2 b_3-a_3 b_2\\
                                 a_1 b_2-a_2 b_1\\
                                 a_3 b_1-a_1 b_3\\
                                \end{array}\right) \, .
\end{equation}


\begin{thebibliography}{0}%
\makeatletter
\providecommand \@ifxundefined [1]{%
 \@ifx{#1\undefined}
}%
\providecommand \@ifnum [1]{%
 \ifnum #1\expandafter \@firstoftwo
 \else \expandafter \@secondoftwo
 \fi
}%
\providecommand \@ifx [1]{%
 \ifx #1\expandafter \@firstoftwo
 \else \expandafter \@secondoftwo
 \fi
}%
\providecommand \natexlab [1]{#1}%
\providecommand \enquote  [1]{``#1''}%
\providecommand \bibnamefont  [1]{#1}%
\providecommand \bibfnamefont [1]{#1}%
\providecommand \citenamefont [1]{#1}%
\providecommand \href@noop [0]{\@secondoftwo}%
\providecommand \href [0]{\begingroup \@sanitize@url \@href}%
\providecommand \@href[1]{\@@startlink{#1}\@@href}%
\providecommand \@@href[1]{\endgroup#1\@@endlink}%
\providecommand \@sanitize@url [0]{\catcode `\\12\catcode `\$12\catcode
  `\&12\catcode `\#12\catcode `\^12\catcode `\_12\catcode `\%12\relax}%
\providecommand \@@startlink[1]{}%
\providecommand \@@endlink[0]{}%
\providecommand \url  [0]{\begingroup\@sanitize@url \@url }%
\providecommand \@url [1]{\endgroup\@href {#1}{\urlprefix }}%
\providecommand \urlprefix  [0]{URL }%
\providecommand \Eprint [0]{\href }%
\providecommand \doibase [0]{http://dx.doi.org/}%
\providecommand \selectlanguage [0]{\@gobble}%
\providecommand \bibinfo  [0]{\@secondoftwo}%
\providecommand \bibfield  [0]{\@secondoftwo}%
\providecommand \translation [1]{[#1]}%
\providecommand \BibitemOpen [0]{}%
\providecommand \bibitemStop [0]{}%
\providecommand \bibitemNoStop [0]{.\EOS\space}%
\providecommand \EOS [0]{\spacefactor3000\relax}%
\providecommand \BibitemShut  [1]{\csname bibitem#1\endcsname}%
\let\auto@bib@innerbib\@empty
\end{thebibliography}%


\begin{thebibliography}{99}

\bibitem{vanderBij:2010nu}
  J.~J.~van der Bij,
  Gen.\  Relativ.\  Gravit.\ 
  [arXiv:1001.3236 [hep-ph]].

\bibitem{Ade:2013zuv}
  P.~A.~R.~Ade {\it et al.} [Planck Collaboration],
  Astron.\ Astrophys.\  {\bf 571} (2014) A16
  [arXiv:1303.5076 [astro-ph.CO]].


\bibitem{Yuan:2014rca}
  Q.~Yuan and B.~Zhang,
  JHEAp {\bf 3-4} (2014) 1
  [arXiv:1404.2318 [astro-ph.HE]].


\bibitem{Gomez-Vargas:2014gra}
  G.~A.~Gómez-Vargas,
  Frascati Phys.\ Ser.\  {\bf 58} (2014) 76
  [arXiv:1410.2376 [hep-ph]].


\bibitem{Calore:2015nua}
  F.~Calore, I.~Cholis and C.~Weniger,
  arXiv:1502.02805 [astro-ph.HE].


\bibitem{O'Leary:2015gfa}
  R.~M.~O'Leary, M.~D.~Kistler, M.~Kerr and J.~Dexter,
  arXiv:1504.02477 [astro-ph.HE].



\bibitem{King:2014nza}
  S.~F.~King, A.~Merle, S.~Morisi, Y.~Shimizu and M.~Tanimoto,
  New J.\ Phys.\  {\bf 16} (2014) 045018
  [arXiv:1402.4271 [hep-ph]].


\bibitem{Varzielas:2011jr}
  I.~de Medeiros Varzielas,
  Phys.\ Lett.\ B {\bf 701} (2011) 597
  [arXiv:1104.2601 [hep-ph]].


\bibitem{Varzielas:2015joa}
  I.~de Medeiros Varzielas, O.~Fischer and V.~Maurer,
  JHEP {\bf 1508} (2015) 080
  [arXiv:1504.03955 [hep-ph]].


\bibitem{Calibbi:2015sfa}
  L.~Calibbi, A.~Crivellin and B.~Zaldívar,
  Phys.\ Rev.\ D {\bf 92} (2015) 1,  016004
  [arXiv:1501.07268 [hep-ph]].


\bibitem{Bishara:2015mha}
  F.~Bishara, A.~Greljo, J.~F.~Kamenik, E.~Stamou and J.~Zupan,
  arXiv:1505.03862 [hep-ph].


\bibitem{Chen:2015jkt}
  M.~C.~Chen, J.~Huang and V.~Takhistov,
  arXiv:1510.04694 [hep-ph].


\bibitem{Ma:2004zd}
  E.~Ma,
  New J.\ Phys.\  {\bf 6} (2004) 104
  [hep-ph/0405152].


\bibitem{deMedeirosVarzielas:2005ax}
  I.~de Medeiros Varzielas and G.~G.~Ross,
  Nucl.\ Phys.\ B {\bf 733} (2006) 31
  [hep-ph/0507176].


\bibitem{deMedeirosVarzielas:2011tp}
  I.~de Medeiros Varzielas, R.~Gonzalez Felipe and H.~Serodio,
  Phys.\ Rev.\ D {\bf 83} (2011) 033007
  [arXiv:1101.0602 [hep-ph]].


\bibitem{Varzielas:2010mp}
  I.~de Medeiros Varzielas and L.~Merlo,
  JHEP {\bf 1102} (2011) 062
  [arXiv:1011.6662 [hep-ph]].


\bibitem{Varzielas:2012ai}
  I.~de Medeiros Varzielas and D.~Pidt,
  JHEP {\bf 1303} (2013) 065
  [arXiv:1211.5370 [hep-ph]].


\bibitem{deMedeirosVarzielas:2011wx}
  I.~de Medeiros Varzielas,
  JHEP {\bf 1201} (2012) 097
  [arXiv:1111.3952 [hep-ph]].


\bibitem{Varzielas:2012ss}
  I.~de Medeiros Varzielas and G.~G.~Ross,
  JHEP {\bf 1212} (2012) 041
  [arXiv:1203.6636 [hep-ph]].


\bibitem{Froggatt:1978nt}
  C.~D.~Froggatt and H.~B.~Nielsen,
  Nucl.\ Phys.\ B {\bf 147} (1979) 277.


\bibitem{Heeck:2014qea}
  J.~Heeck, M.~Holthausen, W.~Rodejohann and Y.~Shimizu,
  Nucl.\ Phys.\ B {\bf 896} (2015) 281
  [arXiv:1412.3671 [hep-ph]].


\bibitem{Alloul:2013bka}
  A.~Alloul, N.~D.~Christensen, C.~Degrande, C.~Duhr and B.~Fuks,
  Comput.\ Phys.\ Commun.\  {\bf 185} (2014) 2250
  [arXiv:1310.1921 [hep-ph]].


\bibitem{Belanger:2014vza}
  G.~Bélanger, F.~Boudjema, A.~Pukhov and A.~Semenov,
  Comput.\ Phys.\ Commun.\  {\bf 192} (2015) 322
  [arXiv:1407.6129 [hep-ph]].


\bibitem{Maurer:2015gva}
  V.~Maurer,
  Comput.\ Phys.\ Commun.\  {\bf 198} (2016) 195
  [arXiv:1503.01073 [cs.MS]].


\bibitem{Antusch:2014woa}
  S.~Antusch and O.~Fischer,
  JHEP {\bf 1410} (2014) 94
  [arXiv:1407.6607 [hep-ph]].


\bibitem{Antusch:2015mia}
  S.~Antusch and O.~Fischer,
  JHEP {\bf 1505} (2015) 053
  [arXiv:1502.05915 [hep-ph]].


\bibitem{Durand:1989zs}
  L.~Durand, J.~M.~Johnson and J.~L.~Lopez,
  Phys.\ Rev.\ Lett.\  {\bf 64} (1990) 1215.


\bibitem{Tsumura:2009yf}
  K.~Tsumura and L.~Velasco-Sevilla,
  Phys.\ Rev.\ D {\bf 81} (2010) 036012
  [arXiv:0911.2149 [hep-ph]].


\bibitem{Berger:2014gga}
  E.~L.~Berger, S.~B.~Giddings, H.~Wang and H.~Zhang,
  Phys.\ Rev.\ D {\bf 90} (2014) 7,  076004
  [arXiv:1406.6054 [hep-ph]].


\bibitem{Babu:1999me}
  K.~S.~Babu and S.~Nandi,
  Phys.\ Rev.\ D {\bf 62} (2000) 033002
  [hep-ph/9907213].


\bibitem{Barger:2012hv}
  V.~Barger, M.~Ishida and W.~Y.~Keung,
  Phys.\ Rev.\ Lett.\  {\bf 108} (2012) 261801
  [arXiv:1203.3456 [hep-ph]].


\bibitem{deFlorian:2015moa}
  D.~de Florian and J.~Mazzitelli,
  JHEP {\bf 1509} (2015) 053
  [arXiv:1505.07122 [hep-ph]].


\bibitem{Gomez-Ceballos:2013zzn}
  M.~Bicer {\it et al.} [TLEP Design Study Working Group Collaboration],
  JHEP {\bf 1401} (2014) 164
  [arXiv:1308.6176 [hep-ex]].


\bibitem{Ruan:2014xxa}
  M.~Ruan,
  arXiv:1411.5606 [hep-ex].


\bibitem{Baer:2013cma}
  H.~Baer {\it et al.},
  arXiv:1306.6352 [hep-ph].


\bibitem{Varzielas:2015iva}
  I.~de Medeiros Varzielas and G.~Hiller,
  JHEP {\bf 1506} (2015) 072
  [arXiv:1503.01084 [hep-ph]].


\bibitem{Adam:2013mnn}
  J.~Adam {\it et al.} [MEG Collaboration],
  Phys.\ Rev.\ Lett.\  {\bf 110} (2013) 201801
  [arXiv:1303.0754 [hep-ex]].


\bibitem{Aaij:2014ora}
  R.~Aaij {\it et al.} [LHCb Collaboration],
  Phys.\ Rev.\ Lett.\  {\bf 113} (2014) 151601
  [arXiv:1406.6482 [hep-ex]].


\bibitem{Hiller:2014yaa}
  G.~Hiller and M.~Schmaltz,
  Phys.\ Rev.\ D {\bf 90} (2014) 054014
  [arXiv:1408.1627 [hep-ph]].


\bibitem{Chatrchyan:2012ep}
  S.~Chatrchyan {\it et al.} [CMS Collaboration],
  JHEP {\bf 1212} (2012) 035
  [arXiv:1209.4533 [hep-ex]].


\bibitem{Chatrchyan:2012zca}
  S.~Chatrchyan {\it et al.} [CMS Collaboration],
  Phys.\ Rev.\ Lett.\  {\bf 110} (2013) 022003
  [arXiv:1209.3489 [hep-ex]].


\bibitem{Aad:2012ux}
  G.~Aad {\it et al.} [ATLAS Collaboration],
  Phys.\ Lett.\ B {\bf 717} (2012) 330
  [arXiv:1205.3130 [hep-ex]].


\bibitem{Aad:2012xca}
  G.~Aad {\it et al.} [ATLAS Collaboration],
  Phys.\ Lett.\ B {\bf 716} (2012) 142
  [arXiv:1205.5764 [hep-ex]].


\bibitem{Aad:2015upn}
  G.~Aad {\it et al.} [ATLAS Collaboration],
  arXiv:1511.05980 [hep-ex].


\bibitem{Kroninger:2015oma}
  K.~Kröninger, A.~B.~Meyer and P.~Uwer,
  arXiv:1506.02800 [hep-ex].

\end{thebibliography}
\end{document}